\documentclass[reprint,5p,sort&compress]{elsarticle}

\usepackage{graphicx,epsfig,float}
\usepackage{bm}
\usepackage{amssymb}
\usepackage{amsmath}
\usepackage[varg]{txfonts}
\usepackage{threeparttable}
\usepackage{makecell}
\usepackage{url}

\renewcommand{\vec}[1]{\bm{\mathrm{#1}}}     

\journal{Physica E: Low-dimensional Systems and Nanostructures}

\begin{document}
\begin{frontmatter}
\title{Robust structural superlubricity of twisted graphene bilayer and  domain walls between commensurate moir\'e pattern domains from first-principles calculations}

\author[nanogune,simune]{Irina V. Lebedeva\corref{cor1}}
\ead{liv\_ira@hotmail.com}
\cortext[cor1]{Corresponding author}
\affiliation[nanogune]{organization={CIC nanoGUNE BRTA},
             addressline={Avenida de Tolosa 76},
             city={San Sebasti\'an},
             postcode={20018},
             country={Spain}}
\affiliation[simune]{organization={Simune Atomistics},
	addressline={Avenida de Tolosa 76},
	city={San Sebasti\'an},
	postcode={20018},
	country={Spain}}

\author[isan]{Andrey M. Popov}
\ead{popov-isan@mail.ru}
\affiliation[isan]{organization={Institute of Spectroscopy of Russian Academy of Sciences},
	addressline={Fizicheskaya str.~5},
	city={Troitsk, Moscow},
	postcode={108840},
	country={Russia}}

\author[kintech]{Yulia G. Polynskaya}
\ead{yupol@kintechlabs.com}
\affiliation[kintech]{organization={Kintech Lab Ltd.},
	addressline={3rd Khoroshevskaya Street 12},
	city={Moscow},
	postcode={123298},
	country={Russia}}

\author[kintech,kurchatov]{Andrey A. Knizhnik}
\ead{knizhnik@kintechlabs.com}
\affiliation[kurchatov]{organization={National Research Centre ``Kurchatov Institute''},
	addressline={Kurchatov Square 1},
	city={Moscow},
	postcode={123298},
	country={Russia}}

\author[bsu]{Sergey A. Vyrko}
\ead{vyrko@bsu.by}
\affiliation[bsu]{organization={Physics Department, Belarusian State University},
	addressline={Nezavisimosti Ave.~4},
	city={Minsk},
	postcode={220030},
	country={Belarus}}

\author[bsu]{Nikolai A. Poklonski}
\ead{poklonski@bsu.by}

\begin{abstract}
Twisted graphene layers exhibit extremely low friction for relative sliding. Nevertheless, previous studies suggest that the area contribution to friction for commensurate moir\'e systems is finite and might restrict macroscopic superlubricity for large layer overlaps.
In this paper, we investigate the potential energy surface (PES) for relative displacement of the layers forming moir\'e patterns (2,1) and (3,1) by accurate density functional theory calculations using the vdW-DF3 functional. The amplitudes of PES corrugations on the order of 0.4 and 0.03 $\mu$eV per atom of one layer, respectively, are obtained. The account of structural relaxation doubles this value for the (2,1) pattern, while causing only minimal changes for the (3,1) pattern.
We show that different from aligned graphene layers, for  moir\'e patterns, PES minima and maxima can switch their positions upon changing the interlayer distance. The PES shape is closely described by the first spatial Fourier harmonics both with and without account of structural relaxation. A barrier for relative rotation of the layers to an incommensurate state that can make superlubricity robust is estimated based on the approximated PES. We also derive a set of measurable physical properties related to interlayer interaction including shear mode frequency, shear modulus and static friction force. Furthermore, we predict that it should be possible to observe domain walls separating commensurate domains, each comprising a large number of moir\'e pattern unit cells, and provide estimates of their characteristics.
\end{abstract}


\end{frontmatter}

\section{Introduction}
\label{sec_intro}
Structural superlubricity is a phenomenon of extremely low friction related to compensation of contributions to friction from surface elementary unit cells having different stackings \cite{Hirano1990, Hirano1991}. Such a phenomenon can be observed for fully incommensurate structures as well as commensurate moir\'e patterns. Graphene is one of 2D materials demonstrating structural superlubricity (see, for example,~\cite{Hod2018} for a review) when the layers are twisted or under tension. If there is no tension applied, superlubricity can be lost via rotation of the layers from the superlubric state to the commensurate ground state \cite{Hirano1990, Verhoeven2004, Dienwiebel2005, Filippov2008, Xu2013, Bonelli2009, vanWijk2013, Guo2007, Shibuta2011, Zhang2015a}. Nevertheless, it has been proposed that superlubricity for twisted bilayer graphene forming a commensurate moir\'e pattern might be robust thanks to the barrier for rotation of the layers to the fully incommensurate state \cite{Minkin2023}. The existence of such a barrier has been confirmed by numerous computational studies of graphene moir\'e patterns \cite{Xu2013, Minkin2023, Belenkov2022, Campanera2007}. Here we investigate structural superlibricity of commensurate moir\'e patterns of graphene using {\it ab initio} calculations.

Superlubricity for graphene was discovered in the experiments for a flake attached to a probe sliding on graphite \cite{Verhoeven2004, Dienwiebel2005, Filippov2008}. In such kind of systems with a small overlap area between the layers moving with respect to each other, the dominant contribution to static friction comes from the edges or rim regions \cite{Verhoeven2004, Filippov2008, Xu2013, Koren2016, Bonelli2009, vanWijk2013, Guo2007, Shibuta2011, Yoon2014, Zhang2015a, Zhang2018, Wang2019a, Zhang2021, Zhang2022, Bai2022, Bai2022a, Tang2023, Yan2024}. There are, however, other sources of friction that can become relevant under different circumstances including: (i) incomplete cancellation of friction forces within full unit cells of commensurate moir\'e patterns (so-called area contribution) \cite{Koren2016, Minkin2023, Minkin2025}, (ii) grain boundaries \cite{Liu2012} and atomic-scale defects \cite{Liu2012,Minkin2021,Minkin2022,Belikov2004, Shibuta2011,Zhang2013a}, (iii) domain walls emerging upon structural relaxation of moir\'e patterns with spatial periods that are much greater than the domain wall width \cite{Mandelli2017, Hod2018, Feng2022}, and (iv) deformation of the layers by the interaction with a substrate \cite{Bai2023}. Obviously diverse atomistic mechanisms are possible not only for static friction as considered in the papers cited above but also for dynamic one \cite{Song2018, Mandelli2017, Wang2019b, Brilliantov2023}. In the present paper we focus on the area contribution (i) to the static friction that, on the one hand, should restrict superlubricity for commensurate moir\'e patterns of bilayer graphene with a large overlap area and, on the other hand, make superlubricity robust by giving rise to the barrier for rotation of the graphene layers to the fully incommensurate state \cite{Minkin2023}.

Friction is determined by the potential energy surface (PES), i.e.~dependence of the potential energy on the relative in-plane displacement of the layers. PESs of commensurate moir\'e patterns of bilayer graphene have been extensively studied using semi-empirical potentials \cite{Minkin2021, Minkin2022, Minkin2023, Minkin2025, Koren2016}. Semi-empirical potentials for interlayer graphene interaction were fitted to the data for aligned graphene layers \cite{Lebedeva2011, Kolmogorov2005} and quantitative results for twisted layers are very different for different potentials \cite{Koren2016, Minkin2021, Minkin2022, Minkin2023}. However, qualitative trends discovered in these works are well justified. It was shown that the PES amplitude, i.e.~the difference in the maximum and minimum potential energies upon relative in-plane displacement of the layers, is much smaller for twisted layers as compared to the aligned ones \cite{Minkin2021, Minkin2022, Minkin2023, Minkin2025} and rapidly decreases upon decreasing the twist angle or equivalently increasing the size of the moir\'e pattern unit cell \cite{Minkin2023, Minkin2025}. At the same time, the PES period also decreases fast upon increasing the size of the moir\'e pattern \cite{Minkin2023, Minkin2025}.

It is clear from these observations that studies of PESs of twisted layers require extremely high energetic and spatial resolution, which makes them very expensive computationally for {\it ab initio} methods. Nevertheless, we are aware of at least two papers in which a non-negligible PES corrugation was observed in density functional theory (DFT) calculations for commensurate moir\'e patterns of bilayer graphene \cite{Song2019, Kabengele2021}. Therefore, the first goal of the present work is to get a more accurate estimate of the PES amplitude for twisted graphene layers by state-of-the-art DFT calculations relying on the known symmetry of PESs for commensurate moir\'e patterns \cite{Minkin2023, Minkin2025}. The effect of structural relaxation on the PES \cite{Minkin2025, Bonelli2009, Zhang2015a, Wang2019, Zhang2018, Zhang2022} in this simulation framework is also addressed.

The second goal of the paper is to relate the computed PES with a set of physical properties that can be measured experimentally. Such experiments would help, on the one hand, to verify the results obtained here and, on the other hand, to improve the theoretical models for description of interaction of graphene layers. Previous studies using semi-empirical potentials showed that the PES shape corresponds to a simple analytical expression that includes only the first spatial Fourier harmonics compatible with symmetries of twisted graphene layers \cite{Minkin2023, Minkin2025}. Such a PES shape seems to be universal for layered 2D systems and has been already confirmed for commensurate graphene and boron nitride layers \cite{Ershova2010, Lebedeva2011, Popov2012, Lebedeva2012, Reguzzoni2012, Lebedev2016, Zhou2015, Lebedev2020}, commensurate double-walled carbon nanotubes \cite{Vucovic2003, Belikov2004, Bichoutskaia2005, Bichoutskaia2009, Popov2009, Popov2012a} and 2D heterostructures \cite{Jung2015, Kumar2015, Lebedev2017}. Here we check the adequacy of the approximation for the PES obtained by DFT calculations and use it to estimate properties of twisted graphene layers related to interlayer interaction, including shear mode frequency, shear modulus and static friction force, that can help to verify the DFT results experimentally. Additionally we propose that, by analogy with aligned graphene bilayers and other 2D materials  \cite{Lin2013, Butz2014, Yankowitz2014, Lebedeva2020, Lebedeva2019a, Lebedeva2019, Yoo2019, Lebedeva2017, Huang2018, SanJose2014, Koshino2013, Lebedev2016, Lebedev2017, Lebedeva2016,  Zhang2025, Enaldiev2020, Soltero2025, Carr2018, Halbertal2021, Kaliteevsk2023}, domain walls separating commensurate domains composed of a large number of moir\'e pattern unit cells can exist in commensurate moir\'e patterns and estimate their characteristics.

The paper is organized in the following way. In section~\ref{sec_model}, we describe the atomic model of twisted graphene bilayer and computational methods. In section~\ref{sec_results}, we present the results of DFT calculations and estimate physical properties of twisted graphene related to interlayer interaction. Section~\ref{sec_conclusions} is devoted to the conclusion.

\section{Model and computational details}
\label{sec_model}

\begin{figure}
   \centering
 \includegraphics{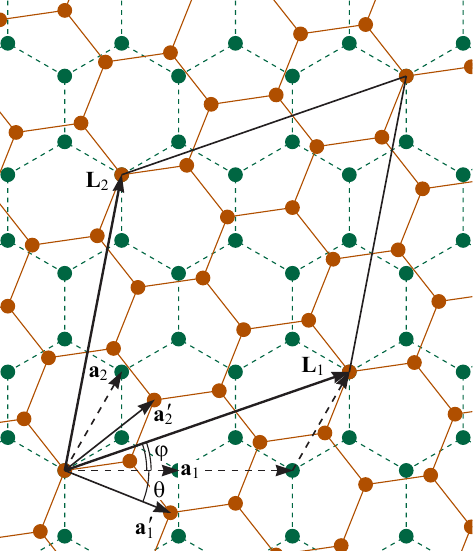}
   \caption{A scheme of the commensurate moir\'e pattern (2,1) formed by twisted graphene layers. Lattice vectors $\vec{a}_1$ and $\vec{a}_2$ of the bottom graphene layer and $\vec{a}_1'$ and $\vec{a}_2'$ of the top layer, lattice vectors $\vec{L}_1$ and $\vec{L}_2$ of the commensurate moir\'e pattern, and twist angle $\theta$ corresponding to relative rotation of the graphene layers are indicated.}
   \label{fig:01}
\end{figure}

Let us briefly discuss the PES for a commensurate moir\'e pattern $(n_1,n_2)$.  To create such a pattern, graphene layers are twisted by the angle $\theta$ determined by
\begin{equation} \label{eq_theta}
   \cos\theta = \frac{n_1^2 + 4n_1n_2 + n_2^2}{2(n_1^2 + n_1n_2 + n_2^2)},
\end{equation}
while the angle between the lattice vector of the bottom graphene layer and the lattice vector of the moir\'e pattern is given by $\varphi = 30^\circ - \theta/2$. The lattice constant of the moir\'e pattern is $L = |\vec{L}_1| = |\vec{L}_2| = a\sqrt{N_\mathrm{c}}$, where $a$ is the graphene lattice constant and $N_\mathrm{c} = n_1^2 + n_1n_2 + n_2^2$ is the number of graphene unit cells per the moir\'e pattern unit cell (Fig.~\ref{fig:01}). 

The PES of aligned layers is described by the first Fourier components, i.e.~contributions from the nearest vertices in the reciprocal space corresponding to reciprocal lattice vectors of magnitude $b = 4\pi/\sqrt{3}a$ \cite{Verhoeven2004, Minkin2023}. When two honeycomb lattices are twisted, the resulting PES should be periodic with respect to translation along each lattice vector of any of them. This means that only overlapping vertices of reciprocal lattices of the twisted layers contribute to the PES Fourier transform. The twisted reciprocal lattices of the honeycomb layers create a moir\'e pattern in the same way as real-space lattices of the layers create the moir\'e pattern in real space. Therefore, (i) the PES for twisted layers is rotated by the same angle  $\varphi$ with respect to the PES for aligned layers as the real-space moir\'e pattern lattice is rotated with respect to the lattice of the bottom graphene layer and (ii) the reciprocal vectors that contribute to the PES are larger by the factor of $L/a=\sqrt{N_\mathrm{c}}$ in magnitude as compared to those for the aligned layers. As a result, the contribution of the first Fourier terms to the PES for twisted layers looks very similar to the PES for aligned layers \cite{Minkin2023, Minkin2025}
\begin{equation} \label{eq_approx}
\delta U(x',y') = U_1\bigg(2\cos{(k'_yy')}\cos{(k'_xx')} +\cos{(2k'_yy')}\bigg)
\end{equation}
but, different from the case of aligned layers, here the $x'$ axis corresponds to the direction of one of the moir\'e pattern lattice vectors, the $y'$ axis to the perpendicular direction ($x' = x \cos\varphi - y \sin\varphi$,  $y' = y \cos\varphi + x \sin\varphi$, where the $x$ axis is aligned along a lattice vector of one of the graphene layers, and the $y$ axis is perpendicular to that vector), while $k'_x = 2\pi\sqrt{N_\mathrm{c}}/a$ and $k'_y = 2\pi\sqrt{N_\mathrm{c}}/\sqrt{3}a$.

As clearly seen from Eq.~(\ref{eq_approx}), the PES period for twisted layers is smaller by the factor of $\sqrt{N_\mathrm{c}}$ than the PES period for aligned layers and smaller by $N_\mathrm{c}$ compared to the moir\'e lattice constant \cite{Minkin2023, Minkin2025}. At the same time, the calculations using the semi-empirical potentials suggest that the PES amplitude for twisted layers is orders of magnitude lower than for the aligned layers \cite{Minkin2021, Minkin2022, Minkin2023, Minkin2025} and decreases exponentially upon increasing $N_\mathrm{c}$ \cite{Minkin2023, Minkin2025}. Therefore, the computational method employed for the PES studies should be able to describe minute changes in energy arising upon small changes in distance with a high accuracy. In the case of plane-wave DFT calculations, this means a high energy cutoff for the plane-wave basis set and a small tolerance for the convergence of self-consistent loop. The cutoff energy $E_\mathrm{cut}$ for the plane-wave basis set is related to the wave vector cutoff $G_\mathrm{cut}$ as $E_\mathrm{cut} = \hbar^2G^2_\mathrm{cut}/2m_e$, where $\hbar$ is the Planck constant and $m_e$ is the electron mass. To resolve the PES for twisted layers, $G_\mathrm{cut}$ should be $\sqrt{N_\mathrm{c}}$ larger than for aligned layers because of the difference in the PES periods for these two systems as discussed above. Correspondingly, the energy cutoff $E_\mathrm{cut}$ used for twisted layers should be at least $N_\mathrm{c}$ times greater than for aligned layers. Since the convergence for aligned layers is reached at the cutoff energy of about 40 Ry \cite{Lebedeva2011}, the cutoff for twisted layers can be estimated as 40 Ry$\cdot N_\mathrm{c}$, which means that the computational cost is huge even for moir\'e patterns with small unit cells. Therefore, in the present paper, we limit our consideration to moir\'e patterns (2,1) and (3,1), which have the smallest unit cells corresponding to $N_\mathrm{c} = 7$ and 13, respectively.

\begin{table*}
	\caption{Properties of aligned bilayer graphene computed using the vdW-DF3 functional (option~1) with and without structure relaxation in comparison with the experimental data: equilibrium interlayer distance $d_\mathrm{eq}$, binding energy in the AB stacking $E_\mathrm{AB}$, amplitude $\Delta U_\mathrm{max} = E_\mathrm{AA} - E_\mathrm{AB}$ of corrugations of the potential energy surface, barrier $\Delta U_\mathrm{b} = E_\mathrm{SP} - E_\mathrm{AB}$ to relative sliding of the layers (all per atom of one layer), and shear mode frequency $f$.}
	\centering
	\renewcommand{\arraystretch}{1.2}
	\setlength{\tabcolsep}{6pt}
	\resizebox{0.8\textwidth}{!}{
		\begin{threeparttable}
			\begin{tabular}{l c c c}
				\hline
				Property & Rigid layers & Out-of-plane relaxation & Experiment \\\hline
				$d_\mathrm{eq}$ (\AA) & 3.3296 & 3.3296 & 3.35 \cite{Brown2012}, 3.3360 $\pm$ 0.0005 \cite{Baskin1955}, 3.355 \cite{Pierson1993}  \\
				$E_\mathrm{AB}$ (meV/atom) & $-52.002$ & $-52.005$ & $-52\pm5$ \cite{Zacharia2004}, $-43\pm5$ \cite{Girifalco1956}, $-35$ ($+15$, $-10$) \cite{Benedict1998}, $-31\pm2$ \cite{Liu2012} \\
				$\Delta U_\mathrm{max}$ (meV/atom)& 15.840 & 10.530 & \\
				$\Delta U_\mathrm{b}$ (meV/atom)& 1.651 & 1.562 & 1.7\tnote{a}~~\cite{Popov2012}, 2.4\tnote{a}~~\cite{Alden2013} \\
				$f$ (cm$^{-1}$)& 23.96 & 23.94 & $28\pm3$ \cite{Boschetto2013}, 32 \cite{Tan2012} \\
				\hline
			\end{tabular}
			\begin{tablenotes}
				\item[a]{An estimate based on the experimental data.}
			\end{tablenotes}
		\end{threeparttable}
	}
	\label{table:aligned}
\end{table*}

The DFT calculations have been carried out using Quantum ESPRESSO \cite{Giannozzi2020, Giannozzi2017, Giannozzi2009, QE}. The exchange-correlation functional vdW-DF3 (option 1) \cite{Chakraborty2020, Thonhauser2015, Thonhauser2007, Berland2015, Langreth2009} including long-range van der Waals interactions is used. This functional adequately predicts the interlayer distance, binding energy and PES amplitude for aligned graphene layers (Table~\ref{table:aligned}, see Ref.~\cite{Lebedeva2016a} for the details of calculation of the shear mode frequency and comparison with other exchange-correlation functionals). The interaction of valence electrons with the core is described using a norm-conserving pseudopotential \cite{Hamann2013, Hamann2017, Setten2018} from the pseudo-dojo database \cite{pseudo-dojo}. One moir\'e pattern unit cell is considered under periodic boundary conditions. The graphene layers are built with the lattice constant $a=2.4660$ \AA, which is the optimal one for the considered functional and pseudopotential according to our calculations and in agreement with previous DFT calculations \cite{Minkin2021, Greshnyakov2019, Lebedeva2011, Popov2012, Lebedeva2016, Lebedeva2016a, Popov2011} and experimental data \cite{Pierson1993}. Accordingly, the lattice constants of the (2,1) and (3,1) moir\'e patterns are $L=6.5245$ \AA{} and 8.8914 \AA, respectively. The Brillouin-zone sampling is performed using the 14$\times$14$\times$1 and 10$\times$10$\times$1 Monkhorst-Pack grids \cite{Monkhorst1976} for the (2,1) and (3,1) moir\'e patterns, respectively. The Gaussian smearing with the width of 0.001 Ry is applied. The self-consistent field iterations are performed till the energy change in consecutive iterations becomes less than $10^{-13}$ Ry.

For the (2,1) moir\'e pattern, we use the energy cutoff for the plane-wave basis set of $E_\mathrm{cut}=400$ Ry. The energy cutoff for the charge density and potential is 2000 Ry. Increasing the cutoffs to 800 and 3200 Ry, respectively, leads only to negligible changes in the PES. The energy cutoffs for the (3,1) pattern are 700 Ry for the basis set and 2800 Ry for the charge density and potential. These cutoffs are chosen to balance the trade-off between the accuracy and computational cost.

The height of the simulation box with periodic boundary conditions should be large enough to avoid the interaction between periodic images. At the same time, in plane-wave calculations, the computational cost grows considerably with increasing the size of the simulation box. To choose the height of the simulation box, we have performed the calculations for rigid layers for the heights of 20 and 30 \AA. For the (2,1) pattern, the effect of the height of the simulation box is negligible. The relative energies of stackings change by less than 0.0013 $\mu$eV per atom of one layer in the case of high-symmetry stackings and less than 0.003 $\mu$eV per atom for other stackings. This is much smaller than the PES amplitude and barrier for relative sliding of the layers. Therefore, for the (2,1) moir\'e pattern, we present the results only for the height of the simulation box of 20 \AA. For the (3,1) pattern, the changes in relative energies of some stackings reach 0.005 $\mu$eV per atom of one layer including stackings close to the symmetric ones. This is comparable to the barrier to relative sliding of the layers. For this reason, the results for rigid layers are presented for the height of the simulation box of 30 \AA. However, since the calculations for this system are rather heavy computationally, the effects of structural relaxation are considered separately for the height of the simulation box of 20 \AA.

A constraint needs to be applied to the layers during the geometry optimization to maintain a given relative in-plane displacement and avoid shifting of the layers to the stacking corresponding to the global energy minimum \cite{Minkin2025}. We consider two types of constraints. To check the effect of out-of-plane relaxation only, in-plane positions are fixed for all atoms. For a more complete description of the relaxation effects, in-plane positions are fixed for one atom of each layer separated by half of the moir\'e pattern lattice vector in each direction. The force tolerance for geometry optimization is $10^{-5}$ Ry/bohr and the energy tolerance is $10^{-9}$ Ry.

The optimal interlayer distance for rigid layers, used in the PES calculation, was determined by performing an energy scan over a range of interlayer distances. Because of the moir\'e pattern inherent symmetry, only a fraction of the PES needs to be computed to characterize its shape and amplitude. An energy profile along a straight line passing through high-symmetry stackings (minima, maxima, and saddle points) is sufficient to reproduce the entire PES \cite{Minkin2025}. To compute the PES profile, we consider the line segment connecting two equivalent extrema and passing through a minimum, maximum, and saddle point. Such a segment has the length $\sqrt{3}l$, where $l=L/N_{\rm c}$ is the length of the PES lattice vector, and is directed along the diagonal of the commensurate moir\'e pattern unit cell. The energy calculations and structure relaxation have been performed for 36 equidistant points within the segment.

\section{Results}
\label{sec_results}
\subsection{DFT calculations of the PES}
\label{subsec_dft}

\begin{table*}
	\caption{Angle $\theta$ of relative rotation of graphene layers, number $N_\mathrm{c}$ of graphene unit cells per the moir\'e pattern unit cell, calculated amplitude $\Delta U_\mathrm{max}$ of corrugations of the potential energy surface, and barrier $\Delta U_\mathrm{b}$ to relative sliding of the layers (both per atom of one layer) for commensurate moir\'e patterns $(n_1,n_2)$ with rigid graphene layers, layers relaxed with constraints on in-plane positions of all atoms and those relaxed with only two constrained atoms in the simulation cell.}
	\renewcommand{\arraystretch}{1.2}
	\setlength{\tabcolsep}{6pt}
	\centering
	\resizebox{\textwidth}{!}{
		\begin{tabular}{*{9}{c}}
			\hline
			&  &  & \multicolumn{2}{c}{Rigid layers} & \multicolumn{2}{c}{Out-of-plane relaxation} & \multicolumn{2}{c}{Relaxation with two atoms constrained} \\
			$(n_1,n_2)$ & $\theta$ (degrees) & $N_\mathrm{c}$  & $\Delta U_\mathrm{max}$ ($\mu$eV/atom) & $\Delta U_\mathrm{b}$ ($\mu$eV/atom) & $\Delta U_\mathrm{max}$ ($\mu$eV/atom) & $\Delta U_\mathrm{b}$ ($\mu$eV/atom) & $\Delta U_\mathrm{max}$ ($\mu$eV/atom) & $\Delta U_\mathrm{b}$ ($\mu$eV/atom)\\\hline
			(2,1) & 21.787 & 7  & 0.347 & 0.048 & 0.577 & 0.060 & 0.765 & 0.078\\
			(3,1) & 32.204 & 13 &  0.027 & $\sim$0.004 & \multicolumn{4}{c}{Expected to be close to the results for rigid layers}  \\
			\hline
		\end{tabular}
	}
	\label{table:U}
\end{table*}

\begin{figure}
	\centering
	\includegraphics{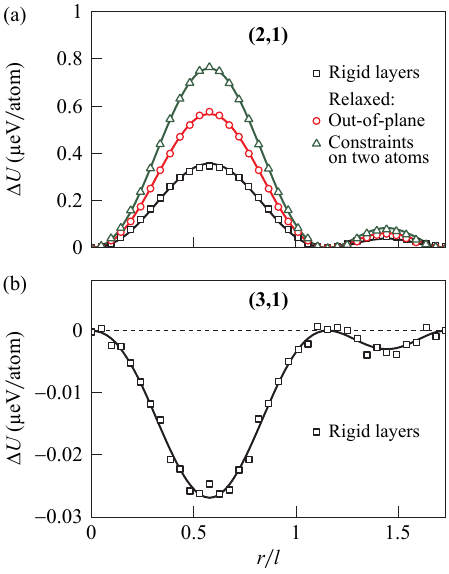}
	\caption{Potential energy change $\Delta U$ (in $\mu$eV per atom of one layer) as a function of the relative displacement $r/l$ of the layers along the diagonal of the moir\'e pattern unit cell for twisted graphene bilayers with commensurate moir\'e patterns (2,1) (a) and (3,1) (b).  The energy is given relative to the PES extrema arranged in a honeycomb lattice. The displacement is given relative to the length $l = L/N_\mathrm{c}$ of the PES lattice vector. For the (2,1) moir\'e pattern (a), the results for rigid layers at the optimal interlayer distance $d_\mathrm{eq} = 3.4022$~\AA{}  (black squares), layers relaxed with constraints on in-plane positions of all atoms (red circles), and those relaxed with only two constrained atoms in the simulation cell (green triangles) are presented. The height of the simulation box is 20 \AA. For the (3,1) moir\'e pattern (b), the results for rigid layers at the optimal interlayer distance $d_\mathrm{eq} = 3.4003$~\AA{} obtained for the height of the simulation box of 30 \AA{} (black squares) are shown. Solid lines correspond to the approximation by the first Fourier harmonics according to Eq.~(\ref{eq_approx}).}\label{fig:pes}
\end{figure}

The PES profiles obtained for the (2,1) and (3,1) moir\'e patterns are presented in Fig.~\ref{fig:pes}. The PES periodicity and shape are consistent with the previous results obtained by calculations with semi-empirical potentials and Eq.~(\ref{eq_approx}) following from symmetry considerations. Two types of the PES in which minima and maxima switch their positions and that correspond to different signs of $U_1$ parameter in Eq.~(\ref{eq_approx}) are possible for twisted graphene layers \cite{Minkin2023, Minkin2025}. According to this classification, the calculated PES for the (2,1) moir\'e pattern is of the second type with maxima and minima in vertices of triangular and honeycomb lattices, respectively, similar to aligned graphene layers. The PES of the (3,1) moir\'e pattern is of the first type and the positions of maxima and minima are inverted as compared to the (2,1) pattern or aligned graphene layers. Note that these PES types are opposite to those predicted for the (2,1) and (3,1) patterns \cite{Minkin2023, Minkin2025} using the Kolmogorov--Crespi potential \cite{Kolmogorov2005}.

The PES amplitudes obtained for rigid layers and with account of structural relaxation (for the (2,1) pattern) are listed in Table~\ref{table:U}. These PES amplitudes are much smaller than the PES corrugations on the order of 60 and 40 $\mu$eV per atom of one layer observed for the (2,1) and (3,1) moir\'e patterns, respectively, in Ref.~\cite{Song2019}. They are also much smaller than the potential energy fluctuations of the maximum amplitude of 7 $\mu$eV per atom of one layer from Ref.~\cite{Kabengele2021}. The reason can be that the parameters used for DFT calculations in those papers are insufficient for accurate PES description for twisted layers. The energy cutoffs were only 30 Ry in Ref.~\cite{Song2019} and 80 Ry in Ref.~\cite{Kabengele2021}, much smaller than those required for such systems (as discussed above). Additionally, the results are sensitive to the pseudopotential choice. Much larger PES fluctuations were observed when employing pseudopotentials different from the norm-conserving pseudopotential used to obtain the results presented in this paper. Other parameters like tolerances for the self-consistent-field loops should also be carefully chosen.

As expected, the PES amplitudes obtained by the DFT calculations (Table~\ref{table:U}, Fig.~\ref{fig:pes}) are also very far from the results generated using the semi-empirical potentials \cite{Minkin2021, Minkin2022, Minkin2023, Minkin2025}. The Kolmogorov--Crespi potential \cite{Kolmogorov2005} gives the PES amplitudes of 87.7 and 20.6 $\mu$eV per atom of one layer for the (2,1) and (3,1) moir\'e patterns with rigid layers, respectively \cite{Minkin2022, Minkin2023, Minkin2025}, i.e.~strongly overestimates the PES amplitude. On the contrary, the Lebedeva potential \cite{Lebedeva2011} gives the vanishing PES amplitude (less than 0.006 $\mu$eV per atom of one layer) even for the (2,1) moir\'e pattern \cite{Minkin2021}. Therefore, common semi-empirical potentials fail to give correct quantitative results. In spite of that, the trends predicted using semi-empirical potentials are physically sound. The decrease of the PES amplitude by more than an order of magnitude for the (3,1) moir\'e pattern compared to (2, 1)  according to the DFT results (Table~\ref{table:U}, Fig.~\ref{fig:pes}) is consistent with the conclusion that the PES amplitude exponentially decreases upon increasing the size of the moir\'e pattern unit cell derived using the Kolmogorov--Crespi potential \cite{Minkin2023, Koren2016, Minkin2025}.

\begin{figure}
	\centering
	\includegraphics{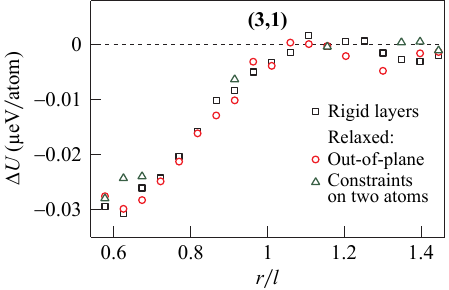}
	\caption{Potential energy change $\Delta U$ (in $\mu$eV per atom of one layer) as a function of the relative displacement $r/l$ of the layers along the diagonal of the moir\'e pattern unit cell for twisted graphene bilayer with the (3,1) moir\'e pattern. The energy is given relative to the PES extrema arranged in a honeycomb lattice. The displacement is given relative to the length $l = L/N_\mathrm{c}$ of the PES lattice vector. The results for rigid layers at the optimal interlayer distance $d_\mathrm{eq} = 3.4003$~\AA{}  (black squares), layers relaxed with constraints on in-plane positions of all atoms (red circles), and those relaxed with only two constrained atoms in the simulation cell (green triangles) are presented. The results are obtained for the height of the simulation box of 20 \AA.}\label{fig:pes_31}
\end{figure}

The DFT calculations show that the total energy of the (2,1) moir\'e pattern is reduced by about 6~$\mu$eV per atom of one layer upon the out-of-plane relaxation and 3~$\mu$eV more when the in-plane relaxation is also taken into account. The PES amplitude, $\Delta U_\mathrm{max}$, however, grows only twice (Table~\ref{table:U}). The out-of-plane relaxation is responsible for 70\% increase of the PES amplitude, while the in-plane relaxation adds 50\% more. For the (3,1) moir\'e pattern, the total energy is reduced by 2.5~$\mu$eV per atom of one layer  upon the out-of-plane relaxation and 2.8~$\mu$eV more when the in-plane relaxation is taken into account as well. The changes in the relative energies of different stackings (corresponding to different $r/l$) for this moir\'e pattern are, nevertheless, comparable to the scatter in the data (Fig.~\ref{fig:pes_31}) and, therefore, in this case we neglect the relaxation effects and consider the PES only for rigid layers (Table~\ref{table:U}). A smaller relative change of the PES amplitude upon relaxation for the (3,1) pattern as compared to the (2,1) one is in agreement with the results obtained using the Kolmogorov--Crepsi potential \cite{Kolmogorov2005} for interlayer interactions and Brenner (REBO-2002) potential \cite{Brenner2002} for intralayer interactions \cite{Minkin2025}.

The equilibrium interlayer distances obtained for twisted layers using the vdW-DF3 functional are close to 3.40 \AA{} (Table~\ref{table:d}). This is greater by 0.07 \AA{} than the spacing for aligned graphene layers (see Table~\ref{table:aligned}) but slightly smaller than typical interlayer distances measured for turbostratic multilayer graphene of 3.41--3.45 \AA{} \cite{Surinlert2022}, 3.42--3.43 \AA{}  \cite{Xu2021}, and 3.435 \AA{} \cite{Kokmat2023}. This is also smaller the interlayer distances around 3.46 \AA{} \cite{Minkin2021, Minkin2023, Minkin2025} computed for twisted layers using the Kolmogorov--Crepsi \cite{Kolmogorov2005} and Lebedeva \cite{Lebedeva2011} potentials. The corrugations $b$ of the layer planes (differences between maximum and minimum $z$-coordinate of atoms within one layer) are considerably larger than the variation $\Delta d$ of the interlayer distance upon shifting the layers with respect to each other (Table~\ref{table:d}).
The minimum and maximum corrugations of the layer planes in relaxed bilayers along the considered displacement path, $b_\mathrm{min}$ and $b_\mathrm{max}$, obtained by the DFT calculations are about twice smaller than those calculated using the semi-empirical potentials for the same patterns \cite{Minkin2025}.

\begin{table*}
	\caption{Calculated equilibrium interlayer distance $d_\mathrm{eq}$ for commensurate moir\'e patterns $(n_1,n_2)$ with rigid graphene layers, as well as average interlayer distance $d_\mathrm{av}$ for layers relaxed with constraints on in-plane positions of all atoms and those relaxed with only two constrained atoms in the simulation cell, difference $\Delta d$ between maximum and minimum interlayer distances for relaxed bilayers, and minimum and maximum corrugations $b$ of the layer plane (difference between maximum and minimum $z$-coordinate of atoms within one layer along the considered displacement path) for relaxed bilayers.}
	\renewcommand{\arraystretch}{1.2}
	\setlength{\tabcolsep}{10pt}
	\centering
	\resizebox{0.9\textwidth}{!}{
		\begin{tabular}{*{10}{c}}
			\hline
			& Rigid layers & \multicolumn{4}{c}{Out-of-plane relaxation} & \multicolumn{4}{c}{Relaxation with two atoms constrained}\\
			$(n_1,n_2)$ & $d_\mathrm{eq}$ (\AA) & $d_\mathrm{av}$ (\AA)& $\Delta d$ (\AA)& $b_\mathrm{min}$ (\AA)& $b_\mathrm{max}$ (\AA)& $d_\mathrm{av}$ (\AA)& $\Delta d$ (\AA)& $b_\mathrm{min}$ (\AA) & $b_\mathrm{max}$ (\AA)\\\hline
			(2,1) & 3.4022 & 3.40214 & 9.3$\cdot10^{-5}$ & 2.16$\cdot10^{-3}$ & 4.26$\cdot10^{-3}$ & 3.40207 &  5.9$\cdot10^{-5}$ & 2.20$\cdot10^{-3}$ & 4.26$\cdot10^{-3}$ \\
			(3,1) & 3.4003 & 3.40025 & 4.8$\cdot10^{-5}$ & 1.51$\cdot10^{-3}$ & 1.67$\cdot10^{-3}$ & 3.40012 & 1.1$\cdot10^{-4}$ & 1.51$\cdot10^{-3}$ & 1.70$\cdot10^{-3}$ \\
			\hline
		\end{tabular}
	}
	\label{table:d}
\end{table*}

\begin{table*}
	\caption{Approximation parameter $U_1$ (per atom of one layer) and relative root-mean-square deviation $\varepsilon$ calculated for commensurate moir\'e patterns $(n_1,n_2)$ with rigid graphene layers, layers relaxed with constraints on in-plane positions of all atoms and those relaxed with only two constrained atoms in the simulation cell for approximation of the potential energy surface using Eq.~(\ref{eq_approx}).}
	\renewcommand{\arraystretch}{1.2}
	\centering
	\setlength{\tabcolsep}{10pt}
	\resizebox{0.75\textwidth}{!}{
		\begin{tabular}{*{7}{c}}
			\hline
			& \multicolumn{2}{c}{Rigid layers} & \multicolumn{2}{c}{Out-of-plane relaxation} & \multicolumn{2}{c}{Relaxation with two atoms constrained} \\
			$(n_1,n_2)$ & $U_1$ ($\mu$eV/atom) & $\varepsilon$ (\%) &  $U_1$ ($\mu$eV/atom)& $\varepsilon$ (\%) &  $U_1$ ($\mu$eV/atom) & $\varepsilon$ (\%) \\
			\hline
			(2,1) & 0.0793 & 2.0 & 0.126 & 0.9 & 0.169 & 0.7 \\
			(3,1) & $-$0.00598 & 3.3 & \multicolumn{4}{c}{Expected to be close to that for rigid layers} \\
			\hline
		\end{tabular}
	}
	\label{table:approx}
\end{table*}

\subsection{PES approximation by the first Fourier harmonics}
\label{subsec_fourier}
As discussed above, the computed PES profiles (Fig.~\ref{fig:pes}) are consistent with the PES shape given by Eq.~(\ref{eq_approx}) following from the Fourier analysis. To evaluate the accuracy of this approximation, we obtained the values of the parameter $U_1$ by minimization of the root-mean-square deviation. These values are presented in Table~\ref{table:approx} along with the relative deviation $\varepsilon$ given by the root-mean-square deviation divided by $\Delta U_\mathrm{max}$.

As seen from Table~\ref{table:approx}, the relative deviation for the (2,1) moir\'e pattern is within 2\%, which means that the PES is closely approximated by the first Fourier harmonics. The accuracy of the approximation is especially high when structural relaxation is taken into account. This is fully consistent with previous observations made using the semi-empirical potentials \cite{Minkin2023, Minkin2025}. For the (3,1) moir\'e pattern, a larger relative deviation of about 3\% is associated with fluctuations in the calculation results due to the finite basis set and accuracy of the pseudopotential used (Fig.~\ref{fig:pes}).

\subsection{Twisted vs. aligned layers}
\label{subsec_vs_aligned}
Despite of apparent similarities in the PES shapes for twisted and aligned graphene layers
(see Eq.~(\ref{eq_approx})), it should be, nevertheless, pointed out  that the behavior of interlayer interaction for twisted layers is in general more complex. This is seen, for example, by the comparison of the PES dependence on the interlayer distance in these two cases.

For aligned graphene layers, the major contribution to the energy variation at a given interlayer distance is provided by the atomic repulsion and the PES amplitude decreases exponentially upon increasing the interlayer distance \cite{Kolmogorov2005, Lebedeva2011, Lebedeva2016a}. The pure  functional of Perdew, Burke and Ernzerhof \cite{Perdew1996} (PBE), although does not show any energy minimum upon changing the interlayer distance, properly describes the PES shape and amplitude at a given interlayer distance \cite{Lebedeva2011, Lebedeva2016a}.

\begin{figure}
	\centering
	\includegraphics{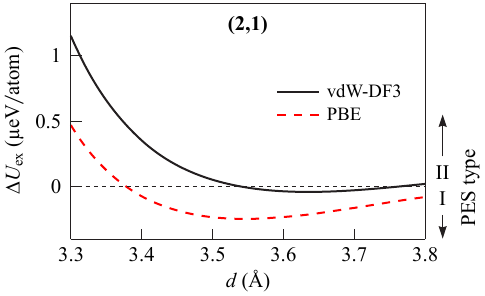}
	\caption{Energy difference $\Delta U_\mathrm{ex}$ between the extrema of the potential energy surface located in vertices of the triangular and honeycomb lattices (in $\mu$eV per atom of one layer, $|\Delta U_\mathrm{ex}| = \Delta U_\mathrm{max}$) for the (2,1) commensurate moir\'e pattern as a function of the interlayer distance $d$ (in \AA) computed using the vdW-DF3 (black line) and PBE (red dashed line) exchange-correlation functionals. The PES types corresponding to different signs of $\Delta U_\mathrm{ex}$ are indicated.}
	\label{fig:extrema}
\end{figure}

\begin{figure}
	\centering
	\includegraphics{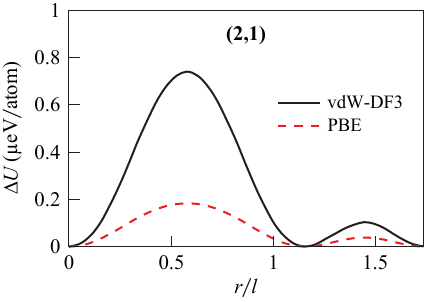}
	\caption{Potential energy change $\Delta U$ (in $\mu$eV per atom of one layer) as a function of the relative displacement $r/l$ of the layers along the diagonal of the moir\'e pattern unit cell for the (2,1) commensurate moir\'e pattern computed using the vdW-DF3 (black line) and PBE (red dashed line) exchange-correlation functionals at the interlayer distance $d=3.34$~\AA. The energy is given relative to the PES minimum. The displacement is given relative to the length $l = L/N_\mathrm{c}$ of the PES lattice vector. }
	\label{fig:pbe_vs_vdwdf}
\end{figure}

Different from the case of the aligned layers, the PES of a commensurate moir\'e pattern can vary in a non-trivial way when the interlayer distance is changed and involve changes in the PES type. To demonstrate this behavior, we consider the energy difference $\Delta U_\mathrm{ex}$ between the PES extrema located in vertices of the triangular and honeycomb lattices (Fig.~\ref{fig:extrema}). The absolute value of this quantity is the same as the PES amplitude $\Delta U_\mathrm{max}$. However, it changes the sign when the PES type is changed. For the PES of the first and second types, $\Delta U_\mathrm{ex}<0$ and $\Delta U_\mathrm{ex}>0$, respectively. Our calculations for the (2,1) moir\'e pattern using the PBE function with no dispersion correction show that $\Delta U_\mathrm{ex}$ goes to zero at the interlayer distance of about 3.4~\AA{} (Fig.~\ref{fig:extrema}) and then changes the sign, that is the PES type changes from second to first. Correspondingly, for the functionals with account of dispersion interactions, the major contribution to the energy variation at distances close to the equilibrium one is provided by the dispersion term (Fig.~\ref{fig:pbe_vs_vdwdf}) and the PES amplitude should be sensitive to the choice of the method for description of such interactions. Furthermore, even with account of dispersion interactions, the PES type change is expected upon increasing the interlayer distance. In the case of the vdW-DF3 functional, the PES becomes extremely smooth at about 3.5~\AA{} and at larger distances, the minima and maxima of the PES switch their positions (Fig.~\ref{fig:extrema}).

\subsection{Physical properties determined by PES}
\label{subsec_physprop}
\begin{table*}
	\caption{Expressions for physical properties of commensurate moir\'e patterns of bilayer graphene related to interlayer interaction.$^a$}
	\renewcommand{\arraystretch}{1.2}
	\setlength{\tabcolsep}{4pt}
	\centering
	\resizebox{0.9\textwidth}{!}{
	\begin{threeparttable}
	\begin{tabular}{*{6}{c}}
		\hline
		& & &  \multicolumn{2}{c}{PES described by the first Fourier harmonics}&  \\
		Notation & Property & General formula & Type I ($U_1<0$) & Type II ($U_1>0$) & Refs. \\
		\hline
		$\Delta U_\mathrm{b}$ & Barrier for relative sliding & & $-4U_1$ & $U_1/2$ &  Eq.~(\ref{eq_approx})\\
		$\Delta U_\mathrm{rot}$ & Barrier for relative rotation & & $-3U_1$ & $3U_1/2$ & \cite{Minkin2023}\\
		$f$  & Shear mode frequency & $\dfrac{1}{2\pi}\sqrt{\dfrac{2}{m_\mathrm{C}}\dfrac{\partial^2 U}{\partial x^2}}$ & $\dfrac{1}{a}\sqrt{\dfrac{-4N_\mathrm{c}U_1}{m_\mathrm{C}}}$ & $\dfrac{1}{a}\sqrt{\dfrac{2N_\mathrm{c}U_1}{m_\mathrm{C}}}$ & \cite{Minkin2023} \\[4ex]
		$C_{44}$ & Shear modulus & $\dfrac{4d}{\sqrt{3}a^2}\dfrac{\partial^2 U}{\partial x^2}$ & $-\dfrac{32\pi^2 d}{\sqrt{3}a^4} N_\mathrm{c}U_1$ & $\dfrac{16\pi^2 d}{\sqrt{3}a^4} N_\mathrm{c}U_1$  & \cite{Minkin2023}\\[4ex]
		$\tau$ & Shear strength & $\dfrac{4}{\sqrt{3}a^2}\max{\left(\dfrac{\partial U}{\partial x_\mathrm{MEP}}\right)}$ & $-\dfrac{16\pi}{\sqrt{3}a^3}\sqrt{N_\mathrm{c}} U_1$ & $\dfrac{6.183}{a^3}\sqrt{N_\mathrm{c}} U_1$  & \cite{Minkin2023}\\[4ex]
		$l_\mathrm{D}(\beta)$ & Domain wall width & $\dfrac{b}{4}\sqrt{\dfrac{\sqrt{3}K(\beta)a^2}{\Delta U_\mathrm{b}}}$ & $\dfrac{a}{8}\sqrt{\dfrac{\sqrt{3}K(\beta)a^2}{ -N_\mathrm{c}U_1}}$ & $\dfrac{a}{2}\sqrt{\dfrac{K(\beta)a^2}{2\sqrt{3}N_\mathrm{c}U_1}}$ & \makecell{\cite{Lebedeva2020, Lebedeva2019a, Lebedeva2019, Lebedev2016, Lebedev2017}\tnote{b}, \\ \cite{Lebedeva2017, Lebedeva2016, Popov2011}\tnote{b}}\\[4ex]
		$W_\mathrm{D}(\beta)$ & Wall formation energy & $\dfrac{2\sqrt{K(\beta)}}{\sqrt[4]{3}a}\int_0^b\sqrt{\Delta U(u)}\:du$ &$\dfrac{8}{\sqrt[4]{3}\pi}\sqrt{\dfrac{-K(\beta)U_1}{N_\mathrm{c}}}$  &$\sqrt{\dfrac{2K(\beta)U_1}{3\sqrt{3}N_\mathrm{c}}}\left(\dfrac{3\sqrt{3}}{\pi}-1 \right)$ &\makecell{\cite{Lebedeva2020, Lebedeva2019a, Lebedeva2019, Lebedev2016, Lebedev2017}\tnote{b}, \\ \cite{Lebedeva2017, Lebedeva2016, Popov2011}\tnote{b}}\\[4ex]
		$\epsilon_\mathrm{u}$ & Critical uniaxial elongation &$\dfrac{W_\mathrm{D}(0)}{kb}$  & $\dfrac{8}{\pi a}\sqrt{\dfrac{-U_1}{\sqrt{3}k(1-\nu^2)}}$ &$\dfrac{1}{a}\sqrt{\dfrac{2U_1}{\sqrt{3}k(1-\nu^2)}}\left(\dfrac{3\sqrt{3}}{\pi}-1 \right)$ &\cite{Popov2011, Lebedeva2016, Lebedev2017}\tnote{b}\\[4ex]
		$\epsilon_\mathrm{b}$ & Critical biaxial elongation &  & &$(1-\nu)\epsilon_\mathrm{u}$ &\cite{Lebedeva2019, Lebedeva2019a, Lebedeva2020}\tnote{b} \\[2ex]
		$F/w$ & Threshold force per unit width & $\dfrac{2\sqrt{K(0)\Delta U_\mathrm{b}}}{\sqrt[4]{3}a}$& $\dfrac{4\sqrt{-K(0)U_1}}{\sqrt[4]{3}a}$& $\dfrac{\sqrt{2K(0)U_1}}{\sqrt[4]{3}a}$& \cite{Popov2011}\tnote{b} \\[2ex]
		\hline
	\end{tabular}
    \begin{tablenotes}
	\item[a]{Here $d_\mathrm{eq}$ is the equilibrium interlayer distance, $a$ is the graphene lattice constant,  $m_\mathrm{C}$ is the mass of a carbon atom, $\partial^2 U/\partial x^2$ is the second-order derivative of the energy per atom of one layer in the energy minimum (independent of the direction because of in-plane isotropy of graphene), $\max{\left(\partial U/\partial x_\mathrm{MEP}\right)}$ is the maximum of the first-order derivative of the energy per atom of one layer with respect to the relative displacement of the layers along the minimum energy path between adjacent energy minima, $b$ is the Burgers vector magnitude (the distance between adjacent energy minima on the PES), $K(\beta)$ equals $k(\cos{\beta}^2 + \sin{\beta}^2(1-\nu)/2)/(1-\nu^2)$ in terms of the elastic constant $k$ under uniaxial stress and the Poisson's ratio $\nu$ and describes the dependence of the elastic constant on the angle $\beta$ between the Burgers vector and normal to the domain wall, $N_\mathrm{c}$ is the number of graphene unit cells per the moir\'e pattern unit cell, and $U_1$ is the parameter of Eq.~(\ref{eq_approx}).}
	\item[b]{Derived here for commensurate moir\'e patterns based on equations presented in previous works.}
	\end{tablenotes}
	\end{threeparttable}
\label{table:prop_exp}
}
\end{table*}

\begin{table*}
	\caption{Barrier $\Delta U_\mathrm{b}$ to relative sliding of the layers (per atom of one layer), barrier $\Delta U_\mathrm{rot}$ for relative rotation of the layers to an incommensurate state (per atom of one layer), shear mode frequencies $f$, shear moduli $C_{44}$, shear strengths $\tau$, widths $l_\mathrm{D}$ and formation energies $W_\mathrm{D}$ of tensile and shear domain walls ($\beta=0$ and $\pi/2$, respectively), critical relative elongations $\epsilon_\mathrm{u}$ and $\epsilon_\mathrm{b}$ for uniaxial and biaxial strains applied to one of the layers for moir\'e patterns (2,1) and (3,1), and threshold force $F_\mathrm{max}/w$ for relative sliding of layers along the line passing through adjacent energy minima per unit overlap width in the limit of large overlaps.}
	\centering
	\renewcommand{\arraystretch}{1.2}
	\setlength{\tabcolsep}{10pt}
	\resizebox{0.8\textwidth}{!}{
	\begin{tabular}{l *{4}{c}}
		\hline
		& \multicolumn{3}{c}{(2,1)} & (3,1) \\
		Property & Rigid layers & Out-of-plane relaxation & Relaxation with two atoms constrained & Rigid layers \\
		\hline
		$\Delta U_\mathrm{b}$ ($\mu$eV/atom) &  0.0397 & 0.0630 & 0.0845 & 0.0239 \\
		$\Delta U_\mathrm{rot}$ ($\mu$eV/atom) & 0.119 &  0.189 & 0.253 & 0.0179 \\
		$f$ (cm$^{-1}$)& 0.404 & 0.510 & 0.590 & 0.214 \\
		$C_{44}$ (Pa) & 7.46$\cdot10^{5}$ & 1.19$\cdot10^{6}$ & 1.59$\cdot10^{6}$ & 2.09$\cdot10^{5}$ \\
		$\tau$ (Pa)& 1.39$\cdot10^{4}$ &  2.20$\cdot10^{4}$ &  2.95$\cdot10^{4}$ &  6.68$\cdot10^{3}$ \\
		$l_\mathrm{D}(0)$ ($\mu$m) & 1.01 & 0.803 & 0.693 & 1.66 \\
		$l_\mathrm{D}(\pi/2)$ ($\mu$m)& 0.650 & 0.516 & 0.446 & 1.06 \\
		$W_\mathrm{D}(0)$ (meV/\AA) & 0.199 & 0.251 & 0.291 & 0.191 \\
		$W_\mathrm{D}(\pi/2)$ (meV/\AA)& 0.128 & 0.162 & 0.187 & 0.123 \\
		$\epsilon_\mathrm{u}$ & $1.79\cdot10^{-5}$ & $2.26\cdot10^{-5}$ & $2.62\cdot10^{-5}$ & $1.36\cdot10^{-5}$\\
		$\epsilon_\mathrm{b}$ &$1.48\cdot10^{-5}$ & $1.87\cdot10^{-5}$ & $2.16\cdot10^{-5}$  & \\
		$F_\mathrm{max}/w$  (N/m) & $9.08\cdot10^{-3}$ & $1.14\cdot10^{-2}$ & $1.32\cdot10^{-2}$ & $7.05\cdot10^{-3}$\\
		\hline
	\end{tabular}
	}
	\label{table:prop}
\end{table*}

The PES determines physical properties of bilayer materials related to interlayer interaction \cite{Minkin2023, Lebedeva2011, Lebedeva2020, Lebedeva2019a, Lebedeva2019, Lebedeva2017, Lebedeva2016a, Lebedeva2016, Lebedeva2012, Lebedev2016, Lebedev2017, Lebedev2020, Popov2011}. On the one hand, this means that the results of the DFT calculations for commensurate moir\'e patterns can be used to predict the properties for these systems. On the other hand, although the results of DFT calculations are expected to be more reliable than those obtained using the semi-empirical potentials, they still need to be verified experimentally. Measurements of any of the physical properties related to interlayer interaction would give a clue on the accuracy of the DFT calculations and would allow to estimate the actual amplitude of PES corrugations.

The barrier $\Delta U_\mathrm{b}$ to relative sliding of the layers corresponding to the energy of the saddle-point stacking with respect to the global energy minimum is in principle available already from the computed PES profiles (Table~\ref{table:U}, Fig.~\ref{fig:pes}). The computational noise, however, makes it difficult to get the value for the (3,1) moir\'e pattern in this way. An estimate  can be obtained based on the approximation by the first Fourier harmonics according to Eq.~(\ref{eq_approx}) (Table~\ref{table:prop_exp}). The barrier $\Delta U_\mathrm{rot}$ for relative rotation of commensurate twisted layers to an incommensurate state, shear strength $\tau$, shear mode frequency $f$ and shear modulus $C_\mathrm{44}$ can be also estimated based on the PES approximation. The expressions for these physical properties derived in Ref.~\cite{Minkin2023} are listed in Table~\ref{table:prop_exp}. The values obtained here from the results of the DFT calculations are given in Table~\ref{table:prop}. Note that they are much smaller than those estimated using the semi-empirical potentials \cite{Minkin2023} because of the orders of magnitude difference in the computed PES amplitudes $\Delta U_\mathrm{max}$. 

Stacking dislocations or domains walls separating commensurate domains in aligned graphene layers is one of the examples of prediction of new phenomena for graphene based on theoretical considerations \cite{Popov2011} that were later confirmed experimentally \cite{Alden2013}. Since then domain walls have been studied for variety of conditions \cite{Lin2013, Butz2014, Yankowitz2014, Lebedeva2020, Lebedeva2019a, Lebedeva2019, Yoo2019, Lebedeva2017, Huang2018, SanJose2014, Koshino2013} and a bunch of materials consisting of 2D layers \cite{Lebedev2016, Lebedev2017, Lebedeva2016,  Zhang2025, Enaldiev2020, Soltero2025, Carr2018, Halbertal2021, Kaliteevsk2023}. By analogy, it can be expected that domain walls separating commensurate domains composed of many moir\'e pattern unit cells are possible for commensurate moir\'e patterns and they can be created upon stretching, pulling or further twisting of layers with respect to each other. Because of the huge difference in PES amplitudes for aligned graphene layers (Table~\ref{table:aligned}) and commensurate moir\'e patterns (Table~\ref{table:U}, Fig.~\ref{fig:pes}), the characteristics of domain walls in these systems should be very different. In the following, we present the corresponding estimates for twisted graphene layers.

According to the two-chain Frenkel--Kontorova model \cite{Lebedeva2020, Lebedeva2019a, Lebedeva2019, Lebedeva2017,Lebedeva2016, Lebedev2016, Lebedev2017, Popov2011, Pokrovskii1978}, the width of domain walls is determined by the elastic properties of the layers, height of the barrier for relative sliding of the layers and Burgers vector, i.e.~the magnitude and direction of relative displacement of the layers required to get from an energy minimum to an adjacent one
\begin{equation} \label{eq_width}
l_\mathrm{D}(\beta) = \frac{b}{2}\sqrt{\frac{K(\beta)}{\Delta V_\mathrm{b}}},
\end{equation}
where $b$ is the magnitude of the Burgers vector equal to the distance between adjacent energy minima on the PES, $K(\beta)$ describes the dependence of the elastic constant on the angle $\beta$ between the Burgers vector and normal to the domain wall, and  $\Delta V_\mathrm{b}$ is the barrier to relative sliding of the layers per unit area. Angles $\beta=0$ and $\pi/2$ correspond to tensile and shear domain walls, respectively. Given that $\Delta V_\mathrm{b} = 4\Delta U_\mathrm{b}/(\sqrt{3}a^2)$, we derive expressions for commensurate moir\'e patterns presented in Table~\ref{table:prop_exp}. Note that the magnitudes $b$ of the Burgers vector are $a/\sqrt{N_\mathrm{c}}$ and $a/\sqrt{3N_\mathrm{c}}$ for the PESs of the first ($U_1<0$) and second type ($U_1>0$), respectively. Using the values of the elastic constant and Poisson's ratio of $k=331\pm1$ J/m$^2$ and $\nu =0.174\pm0.002$, respectively, obtained in the previous DFT calculations \cite{Lebedeva2016}, we estimate that the wall widths for moir\'e patterns (2,1) and (3,1) should be on the order of 1 $\mu$m (Table~\ref{table:prop}). This is two orders of magnitude greater than the domain wall widths for aligned graphene layers \cite{Popov2011, Lebedeva2016, Lebedeva2019, Lebedeva2019a, Lebedeva2020, Alden2013}.

The formation energy of a domain wall in a commensurate moir\'e pattern per unit wall length in the absence of any external load depends not only on the barrier but also on the shape of the potential energy dependence along the minimum energy path between adjacent energy minima \cite{Lebedeva2020, Lebedeva2019a, Lebedeva2019, Lebedeva2017,Lebedeva2016, Lebedev2016, Lebedev2017, Popov2011}
\begin{equation} \label{eq_energy}
W_\mathrm{D}(\beta) = \sqrt{K(\beta)}\int_0^b\sqrt{\Delta V(u)}\:du.
\end{equation}
The expressions derived assuming that the PES is described by the first Fourier harmonics are presented in Table~\ref{table:prop_exp}. Note that the shapes of the potential energy profiles along the minimum energy path for the PESs of the first and second type are slightly different but they are the same as considered in \cite{Popov2011, Lebedev2017} and \cite{Lebedeva2016, Lebedeva2019, Lebedeva2019a, Lebedeva2020}, respectively. The formation energies of domain walls for moir\'e patterns (2,1) and (3,1) are estimated to be on the order of 0.1 meV/\AA{} (Table~\ref{table:prop}), which is three orders of magnitude smaller than for aligned graphene layers \cite{Popov2011, Lebedeva2016, Lebedeva2019, Lebedeva2019a, Lebedeva2020}. 

The total formation energy of a domain wall is proportional to the domain wall length and thus can be significant for large layers in the absence of an external load. The situation changes when strain is applied to one of the layers. At small strains, the bilayer maintains the commensurate pattern. However, upon increasing the strain above some critical value, it becomes energetically favourable to create a region where the layers are incommensurate in order to reduce the elastic energy of the system, that is to create a domain wall. In the limit of infinite bilayer size, the density of domain walls should changes continuously upon increasing the strain above the critical value. Correspondingly, there is a second-order phase
transition from the commensurate to incommensurate phase with the density of domain walls serving as the order parameter \cite{Pokrovskii1978, Popov2011, Lebedeva2020}. The expressions for the critical relative elongation of the layer at which the commensurate-incommensurate phase transition takes place, $\epsilon_\mathrm{u}$ and $\epsilon_\mathrm{b}$, were derived in \cite{Popov2011, Lebedeva2016, Lebedev2017} and \cite{Lebedeva2019, Lebedeva2019a, Lebedeva2020} for uniaxial and biaxial strains, respectively (Table~\ref{table:prop_exp}). In the latter case, only energy minima forming a hexagonal lattice, which corresponds the PES of the second type, were considered. The critical relative elongations estimated for moir\'e patterns (2,1) and (3,1) are on the order of $10^{-5}$, i.e.~two orders of magnitude smaller than for aligned graphene layers. The small critical elongations and formation energies of domain walls in commensurate twisted graphene layers indicate that they should be easily created and observed.

\begin{figure}
	\centering
	\includegraphics{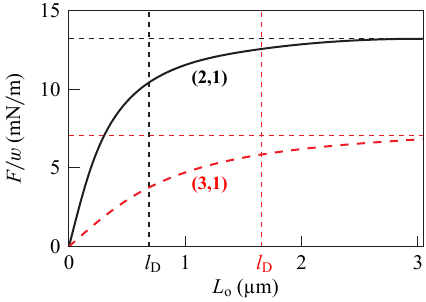}
	\caption{A sketch of the dependence of threshold force $F/w$ (in mN/m) for relative sliding of graphene layers along the line passing through adjacent energy minima on the overlap length $L_\mathrm{o}$ (in $\mu$m). The overlap area is assumed rectangular and the force is given per unit overlap width. The curves for the moir\'e patterns (2,1) and (3,1) are shown by black solid and red dashed lines, respectively. The curve for the (2,1) pattern is drawn based on the results obtained with account of structure relaxation with two atoms constrained. The curve for the (3,1) pattern is drawn based on the results for rigid layers. The widths $l_\mathrm{D}$ of tensile domain walls for these patterns are indicated by vertical dashed lines. Horizontal dashed lines show the asymptotic values corresponding to the formation of domain walls.}
	\label{fig:force}
\end{figure}

Another property that can be measured experimentally to verify the PES characteristics is the threshold force $F$ for relative sliding of graphene layers (or static friction force). At small overlaps of graphene layers, the layers move rigidly with respect to each other when they are pulled away from each other at the opposite ends \cite{Popov2011} and the threshold force is determined by the overlap area $A$ and shear strength  $\tau$ dependent on the PES parameters (Table~\ref{table:prop_exp}), $F=A\tau$ (Fig.~\ref{fig:force}).  Upon increasing the overlap area it becomes energetically favorable to deform the layers by formation of domain walls. In the case of pulling along the line passing through adjacent energy minima (along the moir\'e pattern lattice vector for the PES of the first type and along the diagonal of the moir\'e patter unit cell for the PES of the second type), a tensile domain wall is formed and the threshold force reaches the value of $F_\mathrm{max}=w\sqrt{K(0)\Delta V_\mathrm{b}}$ \cite{Popov2011}, where $w$ is the width of the overlap region perpendicular to the pulling direction, i.e.~$F$ no longer depends on the overlap length along the pulling direction (Fig.~\ref{fig:force}). The expression for this threshold force for two types of the PES and estimates for the (2,1) and (3,1) moir\'e patterns are given in Tables~\ref{table:prop_exp} and \ref{table:prop}, respectively. The crossover between these two modes of static friction occurs when the overlap length is comparable to the width $l_\mathrm{D}(0)$ of tensile domain walls. 

Similar behavior can be expected for the threshold torque for rotation of the layers to the fully incommensurate state. For small overlaps, the threshold torque should be on the order of the relative energy of the incommensurate state $\Delta V_\mathrm{rot} = 4\Delta U_\mathrm{rot}/(a^2\sqrt{3})$ divided by the angle $\delta\varphi \sim a/R$ required to make the system incommensurate, where $R$ is the overlap radius. Therefore, in this case, the threshold torque per unit overlap area grows with the overlap radius as $T\sim 4\Delta U_\mathrm{rot}R/(a^3\sqrt{3})$. For large overlaps, rotation occurs through formation of a network of shear domain walls and the threshold torque per unit overlap area $T \sim \sqrt{K(\pi/2)\Delta V_\mathrm{b}}$ \cite{Lebedeva2019a} no longer depends on the overlap size. The crossover between two modes takes place when the overlap radius $R$ is comparable to the width $l_\mathrm{D}(\pi/2)$ of shear domain walls.

Dynamic friction between macroscopic twisted graphene layers was recently investigated with the help of an atomic force microscope \cite{Brilliantov2023}. We expect that measurements of the threshold force for relative sliding of graphene layers forming a commensurate moir\'e pattern or threshold torque for their rotation  can be performed in a similar manner. More accurate calculations of these quantities can be obtained by using continuum models for the description of layer distortion in combination with Eq.~(\ref{eq_approx}) for the description of interlayer interaction by analogy with \cite{Jung2015, Kumar2015, Kumar2016}. Such an extensive study is, nevertheless, beyond the scope of the present paper.

\section{Conclusions}
\label{sec_conclusions}
We have performed accurate DFT calculations to investigate the PES for relative sliding of the layers forming commensurate moir\'e patterns (2,1) and (3,1). It is shown that the energy cutoff for the plane-wave basis set should be as high as 400 and 700 Ry for the (2,1) and (3,1) patterns, respectively, to get reasonably converged results. The PES amplitudes on the order of 0.4 and 0.03 $\mu$eV per atom of one layer, respectively, are obtained at the optimal interlayer distance. The account of structural relaxation leads to a two-fold increase of this quantity for the (2,1) moir\'e pattern and only small changes for the (3,1) pattern. Both the out-of-plane and in-plane relaxation are important to correctly describe the relaxation effects. 

It is demonstrated that the shape of the PES for the commensurate moir\'e patterns can closely described using the first spatial Fourier harmonics both with and without account of structural relaxation. The accuracy of the PES approximation for the (2,1) pattern, for which the computed potential energy profile is smooth, decreases from 2\% for rigid layers to less than 1\% for relaxed layers. A peculiar behavior of the PES characteristics (amplitude and type) is observed for the (2,1) pattern when the interlayer distance is changed. The PES becomes flat at the interlayer distance of 3.5 \AA, while at larger distances the positions of the PES minima and maxima are switched. 

A set of physical properties including the barriers for relative sliding and rotation of the layers, shear mode frequency, shear modulus, shear strength, and threshold force for sliding (static friction force) are estimated for twisted layers forming moir\'e patterns (2,1) and (3,1) based on the approximated PES. Additionally we believe that, by analogy with bilayers composed of aligned layers of diverse 2D materials \cite{Lin2013, Butz2014, Yankowitz2014, Lebedeva2020, Lebedeva2019a, Lebedeva2019, Yoo2019, Lebedeva2017, Huang2018, SanJose2014, Koshino2013, Lebedev2016, Lebedev2017, Lebedeva2016,  Zhang2025, Enaldiev2020, Soltero2025, Carr2018, Halbertal2021, Kaliteevsk2023}, stretching, pulling or further twisting of the layer in commensurate moir\'e patterns should lead to formation of domain walls separating commensurate domains composed of a large number of moir\'e pattern unit cells. It can be expected that similar to other 2D materials \cite{Yankowitz2014, Yoo2019, Huang2018, Zhang2025, Kaliteevsk2023, SanJose2014, Koshino2013}, such domain walls should also have interesting electronic properties relevant for technological applications.
According to our estimates, the width of domain walls for the commensurate moir\'e patterns lies within the micrometer range, which means that it should be possible to observe domains in the samples of tens of micrometers in size or larger. The formation energy of domain walls and critical relative elongation of the layers required to induce the commensurate-incommensurate phase transition are extremely small for  moir\'e patterns indicating that domain walls should be easily formed in such systems. 

Experimental measurements of any of the above physical properties would help to validate the results on the barrier for relative sliding of twisted graphene layers and refine the quantitative characteristics of the PES. This is important for verification of the hypothesis on the possibility of robust superlubricity in commensurate moir\'e patterns and improvement of {\it ab initio} and other theoretical methods for description of interlayer interaction in 2D materials.

\hfil
\section*{Data availability}
The raw data required to reproduce these findings are available to
download from https://zenodo.org/records/16318186.

\hfil
\section*{Acknowledgments}
A.M.P. and Y.G.P. acknowledge the support by the Russian Science Foundation grant No.~23-42-10010, \url{https://rscf.ru/en/project/23-42-10010/}, for the results described in sections~\ref{subsec_dft}, \ref{subsec_fourier}, and \ref{subsec_vs_aligned}. A.M.P. acknowledges the support by project FFUU-2024-0003 of the Institute of Spectroscopy of the Russian Academy of Sciences for the results described in section~\ref{subsec_physprop}. S.A.V. and N.A.P. acknowledge the support by the Belarusian Republican Foundation for Fundamental Research (Grant No.~F23RNF-049) and by the Belarusian National Research Program ``Convergence-2025''. I.V.L. acknowledges the technical and human support provided by the DIPC Supercomputing Center. 

\bibliographystyle{elsarticle-num} 
\bibliography{twisted_pes.bib}

\begin{thebibliography}{100}
\expandafter\ifx\csname url\endcsname\relax
  \def\url#1{\texttt{#1}}\fi
\expandafter\ifx\csname urlprefix\endcsname\relax\def\urlprefix{URL }\fi
\expandafter\ifx\csname href\endcsname\relax
  \def\href#1#2{#2} \def\path#1{#1}\fi

\bibitem{Hirano1990}
M.~Hirano, K.~Shinjo, Atomistic locking and friction, Phys. Rev. B 41 (1990)
  11837--11851.
\newblock \href {https://doi.org/10.1103/PhysRevB.41.11837}
  {\path{doi:10.1103/PhysRevB.41.11837}}.

\bibitem{Hirano1991}
M.~Hirano, K.~Shinjo, R.~Kaneko, Y.~Murata, Anisotropy of frictional forces in
  muscovite mica, Phys. Rev. Lett. 67 (1991) 2642--2645.
\newblock \href {https://doi.org/10.1103/PhysRevLett.67.2642}
  {\path{doi:10.1103/PhysRevLett.67.2642}}.

\bibitem{Hod2018}
O.~Hod, E.~Meyer, Q.~Zheng, M.~Urbakh, Structural superlubricity and ultralow
  friction across the length scales, Nature 563 (2018) 485--492.
\newblock \href {https://doi.org/10.1038/s41586-018-0704-z}
  {\path{doi:10.1038/s41586-018-0704-z}}.

\bibitem{Verhoeven2004}
G.~S. Verhoeven, M.~Dienwiebel, J.~W.~M. Frenken, Model calculations of
  superlubricity of graphite, Phys. Rev. B 70 (2004) 165418.
\newblock \href {https://doi.org/10.1103/PhysRevB.70.165418}
  {\path{doi:10.1103/PhysRevB.70.165418}}.

\bibitem{Dienwiebel2005}
M.~Dienwiebel, N.~Pradeep, G.~S. Verhoeven, H.~W. Zandbergen, J.~W.~M. Frenken,
  Model experiments of superlubricity of graphite, Surf. Sci. 576 (2005)
  197--211.
\newblock \href {https://doi.org/10.1016/j.susc.2004.12.011}
  {\path{doi:10.1016/j.susc.2004.12.011}}.

\bibitem{Filippov2008}
A.~E. Filippov, M.~Dienwiebel, J.~W.~M. Frenken, J.~Klafter, M.~Urbakh, Torque
  and twist against superlubricity, Phys. Rev. Lett. 100 (2008) 046102.
\newblock \href {https://doi.org/10.1103/PhysRevLett.100.046102}
  {\path{doi:10.1103/PhysRevLett.100.046102}}.

\bibitem{Xu2013}
Z.~Xu, X.~Li, B.~I. Yakobson, F.~Ding, Interaction between graphene layers and
  the mechanisms of graphite's superlubricity and self-retraction, Nanoscale 5
  (2013) 6736--6741.
\newblock \href {https://doi.org/10.1039/c3nr01854g}
  {\path{doi:10.1039/c3nr01854g}}.

\bibitem{Bonelli2009}
F.~Bonelli, N.~Manini, E.~Cadelano, L.~Colombo, Atomistic simulations of the
  sliding friction of graphene flakes, Eur. Phys. J. B 70 (2009) 449--459.
\newblock \href {https://doi.org/10.1140/epjb/e2009-00239-7}
  {\path{doi:10.1140/epjb/e2009-00239-7}}.

\bibitem{vanWijk2013}
M.~M. {van Wijk}, M.~Dienwiebel, J.~W.~M. Frenken, A.~Fasolino, Superlubric to
  stick-slip sliding of incommensurate graphene flakes on graphite, Phys. Rev.
  B 88 (2013) 235423.
\newblock \href {https://doi.org/10.1103/PhysRevB.88.235423}
  {\path{doi:10.1103/PhysRevB.88.235423}}.

\bibitem{Guo2007}
Y.~Guo, W.~Guo, C.~Chen, Modifying atomic-scale friction between two graphene
  sheets: A molecular-force-field study, Phys. Rev. B 76 (2007) 155429.
\newblock \href {https://doi.org/10.1103/PhysRevB.76.155429}
  {\path{doi:10.1103/PhysRevB.76.155429}}.

\bibitem{Shibuta2011}
Y.~Shibuta, J.~A. Elliott, Interaction between two graphene sheets with a
  turbostratic orientational relationship, Chem. Phys. Lett. 512 (2011)
  146--150.
\newblock \href {https://doi.org/10.1016/j.cplett.2011.07.013}
  {\path{doi:10.1016/j.cplett.2011.07.013}}.

\bibitem{Zhang2015a}
H.~Zhang, Z.~Guo, H.~Gao, T.~Chang, Stiffness-dependent interlayer friction of
  graphene, Carbon 94 (2015) 60--66.
\newblock \href {https://doi.org/10.1016/j.carbon.2015.06.024}
  {\path{doi:10.1016/j.carbon.2015.06.024}}.

\bibitem{Minkin2023}
A.~S. Minkin, I.~V. Lebedeva, A.~M. Popov, S.~A. Vyrko, N.~A. Poklonski, Y.~E.
  Lozovik, Interlayer interaction, shear vibrational mode, and tribological
  properties of two-dimensional bilayers with a commensurate moir\'e pattern,
  Phys. Rev. B 108 (2023) 085411.
\newblock \href {https://doi.org/10.1103/PhysRevB.108.085411}
  {\path{doi:10.1103/PhysRevB.108.085411}}.

\bibitem{Belenkov2022}
M.~E. Belenkov, M.~Brzhezinskaya, V.~A. Greshnyakov, E.~A. Belenkov, Modeling
  the structure and interlayer interactions of twisted bilayer graphene,
  Fullerenes, Nanotubes and Carbon Nanostructures 30 (2021) 152--155.
\newblock \href {https://doi.org/10.1080/1536383X.2021.1981295}
  {\path{doi:10.1080/1536383X.2021.1981295}}.

\bibitem{Campanera2007}
J.~M. Campanera, G.~Savini, I.~Suarez-Martinez, M.~I. Heggie, Density
  functional calculations on the intricacies of {M}oir\'e patterns on graphite,
  Phys. Rev. B 75 (2007) 235449.
\newblock \href {https://doi.org/10.1103/PhysRevB.75.235449}
  {\path{doi:10.1103/PhysRevB.75.235449}}.

\bibitem{Koren2016}
E.~Koren, U.~Duerig, Moir\'e scaling of the sliding force in twisted bilayer
  graphene, Phys. Rev. B 94 (2016) 045401.
\newblock \href {https://doi.org/10.1103/PhysRevB.94.045401}
  {\path{doi:10.1103/PhysRevB.94.045401}}.

\bibitem{Yoon2014}
H.~M. Yoon, S.~Kondaraju, J.~S. Lee, Molecular dynamics simulations of the
  friction experienced by graphene flakes in rotational motion, Tribol. Int. 70
  (2014) 170--178.
\newblock \href {https://doi.org/10.1016/j.triboint.2013.10.005}
  {\path{doi:10.1016/j.triboint.2013.10.005}}.

\bibitem{Zhang2018}
H.~Zhang, T.~Chang, Edge orientation dependent nanoscale friction, Nanoscale 10
  (2018) 2447--2453.
\newblock \href {https://doi.org/10.1039/C7NR07839K}
  {\path{doi:10.1039/C7NR07839K}}.

\bibitem{Wang2019a}
W.~Wang, J.~Shen, Q.-C. He, Microscale superlubricity of graphite under various
  twist angles, Phys. Rev. B 99 (2019) 054103.
\newblock \href {https://doi.org/10.1103/PhysRevB.99.054103}
  {\path{doi:10.1103/PhysRevB.99.054103}}.

\bibitem{Zhang2021}
H.~Zhang, Y.~Li, J.~Qu, J.~Zhang, Edge length-dependent interlayer friction of
  graphene, RSC Adv. 11 (2021) 328--334.
\newblock \href {https://doi.org/10.1039/D0RA08457C}
  {\path{doi:10.1039/D0RA08457C}}.

\bibitem{Zhang2022}
H.~Zhang, J.~Qu, Z.~Guo, L.~Huang, Q.~Xie, Negative area-dependent nanoscale
  friction of annular graphene sheets, AIP Adv. 12 (2022) 115312.
\newblock \href {https://doi.org/10.1063/5.0117212}
  {\path{doi:10.1063/5.0117212}}.

\bibitem{Bai2022}
H.~Bai, H.~Bao, Y.~Li, H.~Xu, S.~Li, F.~Ma, One-dimensional strain solitons
  manipulated superlubricity on graphene interface, J. Phys. Chem. Lett. 13
  (2022) 7261--7268.
\newblock \href {https://doi.org/10.1021/acs.jpclett.2c02066}
  {\path{doi:10.1021/acs.jpclett.2c02066}}.

\bibitem{Bai2022a}
H.~Bai, H.~Bao, Y.~Li, H.~Xu, S.~Li, F.~Ma, Moir\'e pattern based universal
  rules governing interfacial superlubricity: A case of graphene, Carbon 191
  (2022) 28--35.
\newblock \href {https://doi.org/10.1016/j.carbon.2022.01.047}
  {\path{doi:10.1016/j.carbon.2022.01.047}}.

\bibitem{Tang2023}
K.~Tang, G.~Ru, W.~Qi, W.~Liu, Moir\'e pattern effect on sliding friction of
  two-dimensional materials, Tribol. Int. 180 (2023) 108288.
\newblock \href
  {https://doi.org/https://doi.org/10.1016/j.triboint.2023.108288}
  {\path{doi:https://doi.org/10.1016/j.triboint.2023.108288}}.

\bibitem{Yan2024}
W.~Yan, X.~Gao, W.~Ouyang, Z.~Liu, O.~Hod, M.~Urbakh, Shape-dependent friction
  scaling laws in twisted layered material interfaces, J. Mech. Phys. Solids
  185 (2024) 105555.
\newblock \href {https://doi.org/10.1016/j.jmps.2024.105555}
  {\path{doi:10.1016/j.jmps.2024.105555}}.

\bibitem{Minkin2025}
A.~S. Minkin, I.~V. Lebedeva, A.~M. Popov, S.~A. Vyrko, N.~A. Poklonski, Y.~E.
  Lozovik, Restriction of macroscopic structural superlubricity due to
  structure relaxation by the example of twisted graphene bilayer, Phys. Rev.
  Mater. 9 (2025) 024002.
\newblock \href {https://doi.org/10.1103/PhysRevMaterials.9.024002}
  {\path{doi:10.1103/PhysRevMaterials.9.024002}}.

\bibitem{Liu2012}
Z.~Liu, J.~Yang, F.~Grey, J.~Z. Liu, Y.~Liu, Y.~Wang, Y.~Yang, Y.~Cheng,
  Q.~Zheng, Observation of microscale superlubricity in graphite, Phys. Rev.
  Lett. 108 (2012) 205503.
\newblock \href {https://doi.org/10.1103/PhysRevLett.108.205503}
  {\path{doi:10.1103/PhysRevLett.108.205503}}.

\bibitem{Minkin2021}
A.~S. Minkin, I.~V. Lebedeva, A.~M. Popov, A.~A. Knizhnik, Atomic-scale defects
  restricting structural superlubricity: \textit{Ab initio} study on the
  example of the twisted graphene bilayer, Phys. Rev. B 104 (2021) 075444.
\newblock \href {https://doi.org/10.1103/PhysRevB.104.075444}
  {\path{doi:10.1103/PhysRevB.104.075444}}.

\bibitem{Minkin2022}
A.~S. Minkin, I.~V. Lebedeva, A.~M. Popov, A.~A. Knizhnik, Simulation of
  tribological properties of a graphene bilayer with twisted layers,
  Nanotechnol. Russia 17 (2022) 472--476.
\newblock \href {https://doi.org/10.1134/S2635167622040176}
  {\path{doi:10.1134/S2635167622040176}}.

\bibitem{Belikov2004}
A.~V. Belikov, Y.~E. Lozovik, A.~G. Nikolaev, A.~M. Popov, Double-wall
  nanotubes: classification and barriers to walls relative rotation, sliding
  and screwlike motion, Chem. Phys. Lett. 385 (2004) 72--78.
\newblock \href {https://doi.org/10.1016/j.cplett.2003.12.049}
  {\path{doi:10.1016/j.cplett.2003.12.049}}.

\bibitem{Zhang2013a}
R.~Zhang, Z.~Ning, Y.~Zhang, Q.~Zheng, Q.~Chen, H.~Xie, Q.~Zhang, W.~Qian,
  F.~Wei, Superlubricity in centimetres-long double-walled carbon nanotubes
  under ambient conditions, Nat. Nanotechnol. 8 (2013) 912--916.
\newblock \href {https://doi.org/10.1038/nnano.2013.217}
  {\path{doi:10.1038/nnano.2013.217}}.

\bibitem{Mandelli2017}
D.~Mandelli, I.~Leven, O.~Hod, M.~Urbakh, Sliding friction of
  graphene/hexagonal-boron nitride heterojunctions: a route to robust
  superlubricity, Sci. Rep. 7 (2017) 10851.
\newblock \href {https://doi.org/10.1038/s41598-017-10522-8}
  {\path{doi:10.1038/s41598-017-10522-8}}.

\bibitem{Feng2022}
S.~Feng, Z.~Xu, Robustness of structural superlubricity beyond rigid models,
  Friction 10 (2022) 1382--1392.
\newblock \href {https://doi.org/10.1007/s40544-021-0548-7}
  {\path{doi:10.1007/s40544-021-0548-7}}.

\bibitem{Bai2023}
H.~Bai, G.~Zou, H.~Bao, S.~Li, F.~Ma, H.~Gao, Deformation coupled moir\'e
  mapping of superlubricity in graphene, ACS Nano 17 (2023) 12594--12602.
\newblock \href {https://doi.org/10.1021/acsnano.3c02915}
  {\path{doi:10.1021/acsnano.3c02915}}.

\bibitem{Song2018}
Y.~Song, D.~Mandelli, O.~Hod, M.~Urbakh, M.~Ma, Q.~Zheng, Robust microscale
  superlubricity in graphite/hexagonal boron nitride layered heterojunctions,
  Nat. Mater. 17 (2018) 894--899.
\newblock \href {https://doi.org/10.1038/s41563-018-0144-z}
  {\path{doi:10.1038/s41563-018-0144-z}}.

\bibitem{Wang2019b}
J.~Wang, W.~Cao, Y.~Song, C.~Qu, Q.~Zheng, M.~Ma, Generalized scaling law of
  structural superlubricity, Nano Letters 19~(11) (2019) 7735--7741.
\newblock \href {https://doi.org/10.1021/acs.nanolett.9b02656}
  {\path{doi:10.1021/acs.nanolett.9b02656}}.

\bibitem{Brilliantov2023}
N.~V. Brilliantov, A.~A. Tsukanov, A.~K. Grebenko, A.~G. Nasibulin, I.~A.
  Ostanin, Atomistic mechanism of friction-force independence on the normal
  load and other friction laws for dynamic structural superlubricity, Phys.
  Rev. Lett. 131 (2023) 266201.
\newblock \href {https://doi.org/10.1103/PhysRevLett.131.266201}
  {\path{doi:10.1103/PhysRevLett.131.266201}}.

\bibitem{Lebedeva2011}
I.~V. Lebedeva, A.~A. Knizhnik, A.~M. Popov, Y.~E. Lozovik, B.~V. Potapkin,
  Interlayer interaction and relative vibrations of bilayer graphene, Phys.
  Chem. Chem. Phys. 13 (2011) 5687--5695.
\newblock \href {https://doi.org/10.1039/c0cp02614j}
  {\path{doi:10.1039/c0cp02614j}}.

\bibitem{Kolmogorov2005}
A.~N. Kolmogorov, V.~H. Crespi, Registry-dependent interlayer potential for
  graphitic systems, Phys. Rev. B 71 (2005) 235415.
\newblock \href {https://doi.org/10.1103/PhysRevB.71.235415}
  {\path{doi:10.1103/PhysRevB.71.235415}}.

\bibitem{Song2019}
H.-Q. Song, Z.~Liu, D.-B. Zhang, Interlayer vibration of twisted bilayer
  graphene: A first-principles study, Phys. Lett. A 383 (2019) 2628--2632.
\newblock \href {https://doi.org/10.1016/j.physleta.2019.05.025}
  {\path{doi:10.1016/j.physleta.2019.05.025}}.

\bibitem{Kabengele2021}
T.~Kabengele, E.~R. Johnson, Theoretical modeling of structural superlubricity
  in rotated bilayer graphene{,} hexagonal boron nitride{,} molybdenum
  disulfide{,} and blue phosphorene, Nanoscale 13 (2021) 14399--14407.
\newblock \href {https://doi.org/10.1039/D1NR03001A}
  {\path{doi:10.1039/D1NR03001A}}.

\bibitem{Wang2019}
K.~Wang, W.~Ouyang, W.~Cao, M.~Ma, Q.~Zheng, Robust superlubricity by strain
  engineering, Nanoscale 11 (2019) 2186--2193.
\newblock \href {https://doi.org/10.1039/c8nr07963c}
  {\path{doi:10.1039/c8nr07963c}}.

\bibitem{Ershova2010}
O.~V. Ershova, T.~C. Lillestolen, E.~Bichoutskaia, Study of polycyclic aromatic
  hydrocarbons adsorbed on graphene using density functional theory with
  empirical dispersion correction, Phys. Chem. Chem. Phys. 12 (2010)
  6483--6491.
\newblock \href {https://doi.org/10.1039/C000370K}
  {\path{doi:10.1039/C000370K}}.

\bibitem{Popov2012}
A.~M. Popov, I.~V. Lebedeva, A.~A. Knizhnik, Y.~E. Lozovik, B.~V. Potapkin,
  Barriers to motion and rotation of graphene layers based on measurements of
  shear mode frequencies, Chem. Phys. Lett. 536 (2012) 82--86.
\newblock \href {https://doi.org/10.1016/j.cplett.2012.03.082}
  {\path{doi:10.1016/j.cplett.2012.03.082}}.

\bibitem{Lebedeva2012}
I.~V. Lebedeva, A.~A. Knizhnik, A.~M. Popov, Y.~E. Lozovik, B.~V. Potapkin,
  Modeling of graphene-based {NEMS}, Physica E 44 (2012) 949--954.
\newblock \href {https://doi.org/10.1016/j.physe.2011.07.018}
  {\path{doi:10.1016/j.physe.2011.07.018}}.

\bibitem{Reguzzoni2012}
M.~Reguzzoni, A.~Fasolino, E.~Molinari, M.~C. Righi, Potential energy surface
  for graphene on graphene: \textit{Ab initio} derivation, analytical
  description, and microscopic interpretation, Phys. Rev. B 86 (2012) 245434.
\newblock \href {https://doi.org/10.1103/PhysRevB.86.245434}
  {\path{doi:10.1103/PhysRevB.86.245434}}.

\bibitem{Lebedev2016}
A.~V. Lebedev, I.~V. Lebedeva, A.~A. Knizhnik, A.~M. Popov, Interlayer
  interaction and related properties of bilayer hexagonal boron nitride:
  \textit{ab initio} study, RSC Adv. 6 (2016) 6423--6435.
\newblock \href {https://doi.org/10.1039/C5RA20882C}
  {\path{doi:10.1039/C5RA20882C}}.

\bibitem{Zhou2015}
S.~Zhou, J.~Han, S.~Dai, J.~Sun, D.~J. Srolovitz, Van der {W}aals bilayer
  energetics: Generalized stacking-fault energy of graphene, boron nitride, and
  graphene/boron nitride bilayers, Phys. Rev. B 92 (2015) 155438.
\newblock \href {https://doi.org/10.1103/PhysRevB.92.155438}
  {\path{doi:10.1103/PhysRevB.92.155438}}.

\bibitem{Lebedev2020}
A.~V. Lebedev, I.~V. Lebedeva, A.~M. Popov, A.~A. Knizhnik, N.~A. Poklonski,
  S.~A. Vyrko, Universal description of potential energy surface of interlayer
  interaction in two-dimensional materials by first spatial {F}ourier
  harmonics, Phys. Rev. B 102 (2020) 045418.
\newblock \href {https://doi.org/10.1103/PhysRevB.102.045418}
  {\path{doi:10.1103/PhysRevB.102.045418}}.

\bibitem{Vucovic2003}
T.~Vukovi\'c, M.~Damnjanovi\'c, I.~Milo\v{s}evi\'c, Interaction between layers
  of the multi-wall carbon nanotubes, Physica E 16 (2003) 259--268.
\newblock \href {https://doi.org/10.1016/S1386-9477(02)00685-9}
  {\path{doi:10.1016/S1386-9477(02)00685-9}}.

\bibitem{Bichoutskaia2005}
E.~Bichoutskaia, A.~M. Popov, A.~El-Barbary, M.~I. Heggie, Y.~E. Lozovik,
  \textit{Ab initio} study of relative motion of walls in carbon nanotubes,
  Phys. Rev. B 71 (2005) 113403.
\newblock \href {https://doi.org/10.1103/PhysRevB.71.113403}
  {\path{doi:10.1103/PhysRevB.71.113403}}.

\bibitem{Bichoutskaia2009}
E.~Bichoutskaia, A.~M. Popov, Y.~E. Lozovik, O.~V. Ershova, I.~V. Lebedeva,
  A.~A. Knizhnik, Modeling of an ultrahigh-frequency resonator based on the
  relative vibrations of carbon nanotubes, Phys. Rev. B 80 (2009) 165427.
\newblock \href {https://doi.org/10.1103/PhysRevB.80.165427}
  {\path{doi:10.1103/PhysRevB.80.165427}}.

\bibitem{Popov2009}
A.~M. Popov, Y.~E. Lozovik, A.~S. Sobennikov, A.~A. Knizhnik, Nanomechanical
  properties and phase transitions in a double-walled (5,5)@(10,10) carbon
  nanotube: {\it ab initio} calculations, JETP 108 (2009) 621--628.
\newblock \href {https://doi.org/10.1134/S1063776109040104}
  {\path{doi:10.1134/S1063776109040104}}.

\bibitem{Popov2012a}
A.~M. Popov, I.~V. Lebedeva, A.~A. Knizhnik, Nanotube-based scanning rotational
  microscope, Appl. Phys. Lett. 100 (2012) 173101.
\newblock \href {https://doi.org/10.1063/1.4705430}
  {\path{doi:10.1063/1.4705430}}.

\bibitem{Jung2015}
J.~Jung, A.~M. DaSilva, A.~H. MacDonald, S.~Adam, Origin of band gaps in
  graphene on hexagonal boron nitride, Nat. Commun. 6 (2015) 6308.
\newblock \href {https://doi.org/10.1038/ncomms7308}
  {\path{doi:10.1038/ncomms7308}}.

\bibitem{Kumar2015}
H.~Kumar, D.~Er, L.~Dong, J.~Li, V.~B. Shenoy, Elastic deformations in {2D} van
  der {W}aals heterostructures and their impact on optoelectronic properties:
  Predictions from a multiscale computational approach, Sci. Rep. 5 (2015)
  10872.
\newblock \href {https://doi.org/10.1038/srep10872}
  {\path{doi:10.1038/srep10872}}.

\bibitem{Lebedev2017}
A.~V. Lebedev, I.~V. Lebedeva, A.~M. Popov, A.~A. Knizhnik, Stacking in
  incommensurate graphene/hexagonal-boron-nitride heterostructures based on
  \textit{ab initio} study of interlayer interaction, Phys. Rev. B 96 (2017)
  085432.
\newblock \href {https://doi.org/10.1103/PhysRevB.96.085432}
  {\path{doi:10.1103/PhysRevB.96.085432}}.

\bibitem{Lin2013}
J.~Lin, W.~Fang, W.~Zhou, A.~R. Lupini, J.~C. Idrobo, J.~Kong, S.~J. Pennycook,
  S.~T. Pantelides, {AC/AB} stacking boundaries in bilayer graphene, Nano
  Letters 13 (2013) 3262--3268.
\newblock \href {https://doi.org/10.1021/nl4013979}
  {\path{doi:10.1021/nl4013979}}.

\bibitem{Butz2014}
B.~Butz, C.~Dolle, F.~Niekiel, K.~Weber, D.~Waldmann, H.~B. Weber, B.~Meyer,
  E.~Spiecker, Dislocations in bilayer graphene, Nature 505 (2014) 533--537.
\newblock \href {https://doi.org/10.1038/nature12780}
  {\path{doi:10.1038/nature12780}}.

\bibitem{Yankowitz2014}
M.~Yankowitz, J.~{\^I-J}. Wang, A.~G. Birdwell, Y.-A. Chen, K.~Watanabe,
  T.~Taniguchi, P.~Jacquod, P.~San-Jose, P.~Jarillo-Herrero, B.~J. LeRoy,
  Electric field control of soliton motion and stacking in trilayer graphene,
  Nature Materials 13~(8) (2014) 786--789.
\newblock \href {https://doi.org/10.1038/nmat3965}
  {\path{doi:10.1038/nmat3965}}.

\bibitem{Lebedeva2020}
I.~V. Lebedeva, A.~M. Popov, Two phases with different domain wall networks and
  a reentrant phase transition in bilayer graphene under strain, Phys. Rev.
  Lett. 124 (2020) 116101.
\newblock \href {https://doi.org/10.1103/PhysRevLett.124.116101}
  {\path{doi:10.1103/PhysRevLett.124.116101}}.

\bibitem{Lebedeva2019a}
I.~V. Lebedeva, A.~M. Popov, Energetics and structure of domain wall networks
  in minimally twisted bilayer graphene under strain, J. Phys. Chem. C 124
  (2019) 2120--2130.
\newblock \href {https://doi.org/10.1021/acs.jpcc.9b08306}
  {\path{doi:10.1021/acs.jpcc.9b08306}}.

\bibitem{Lebedeva2019}
I.~V. Lebedeva, A.~M. Popov, Commensurate-incommensurate phase transition and a
  network of domain walls in bilayer graphene with a biaxially stretched layer,
  Phys. Rev. B 99 (2019) 195448.
\newblock \href {https://doi.org/10.1103/PhysRevB.99.195448}
  {\path{doi:10.1103/PhysRevB.99.195448}}.

\bibitem{Yoo2019}
H.~Yoo, R.~Engelke, S.~Carr, S.~Fang, K.~Zhang, P.~Cazeaux, S.~H. Sung,
  R.~Hovden, A.~W. Tsen, T.~Taniguchi, K.~Watanabe, G.-C. Yi, M.~Kim,
  M.~Luskin, E.~B. Tadmor, E.~Kaxiras, P.~Kim, Atomic and electronic
  reconstruction at the van der {Waals} interface in twisted bilayer graphene,
  Nat. Mater. 18 (2019) 448--453.
\newblock \href {https://doi.org/https://doi.org/10.1038/s41563-019-0346-z}
  {\path{doi:https://doi.org/10.1038/s41563-019-0346-z}}.

\bibitem{Lebedeva2017}
I.~V. Lebedeva, A.~A. Knizhnik, A.~M. Popov, Edge stacking dislocations in
  two-dimensional bilayers with a small lattice mismatch, Physica E 90 (2017)
  49--54.
\newblock \href {https://doi.org/10.1016/j.physe.2017.03.008}
  {\path{doi:10.1016/j.physe.2017.03.008}}.

\bibitem{Huang2018}
S.~Huang, K.~Kim, D.~K. Efimkin, T.~Lovorn, T.~Taniguchi, K.~Watanabe, A.~H.
  MacDonald, E.~Tutuc, B.~J. LeRoy, Topologically protected helical states in
  minimally twisted bilayer graphene, Phys. Rev. Lett. 121 (2018) 037702.
\newblock \href {https://doi.org/10.1103/PhysRevLett.121.037702}
  {\path{doi:10.1103/PhysRevLett.121.037702}}.

\bibitem{SanJose2014}
P.~San-Jose, R.~V. Gorbachev, A.~K. Geim, K.~S. Novoselov, F.~Guinea, Stacking
  boundaries and transport in bilayer graphene, Nano Letters 14 (2014)
  2052--2057.
\newblock \href {https://doi.org/10.1021/nl500230a}
  {\path{doi:10.1021/nl500230a}}.

\bibitem{Koshino2013}
M.~Koshino, Electronic transmission through {$AB$--$BA$} domain boundary in
  bilayer graphene, Phys. Rev. B 88 (2013) 115409.
\newblock \href {https://doi.org/10.1103/PhysRevB.88.115409}
  {\path{doi:10.1103/PhysRevB.88.115409}}.

\bibitem{Lebedeva2016}
I.~V. Lebedeva, A.~V. Lebedev, A.~M. Popov, A.~A. Knizhnik, Dislocations in
  stacking and commensurate-incommensurate phase transition in bilayer graphene
  and hexagonal boron nitride, Phys. Rev. B 93 (2016) 235414.
\newblock \href {https://doi.org/10.1103/PhysRevB.93.235414}
  {\path{doi:10.1103/PhysRevB.93.235414}}.

\bibitem{Zhang2025}
B.~Zhang, L.~He, S.~Zhang, Y.~Ni, Universal state transition from soliton
  network to soliton-free incommensurate stacking in reconstructed van der
  {Waals} bilayers, 2D Materials 12 (2025) 025007.
\newblock \href {https://doi.org/10.1088/2053-1583/ada621}
  {\path{doi:10.1088/2053-1583/ada621}}.

\bibitem{Enaldiev2020}
V.~V. Enaldiev, V.~Z\'olyomi, C.~Yelgel, S.~J. Magorrian, V.~I. Fal'ko,
  Stacking domains and dislocation networks in marginally twisted bilayers of
  transition metal dichalcogenides, Phys. Rev. Lett. 124 (2020) 206101.
\newblock \href {https://doi.org/10.1103/PhysRevLett.124.206101}
  {\path{doi:10.1103/PhysRevLett.124.206101}}.

\bibitem{Soltero2025}
I.~Soltero, V.~I. Fal’ko, Interlayer dislocations in multilayer and bulk
  {MoS$_2$}, 2D Materials 12 (2025) 025003.
\newblock \href {https://doi.org/10.1088/2053-1583/ad9f7d}
  {\path{doi:10.1088/2053-1583/ad9f7d}}.

\bibitem{Carr2018}
S.~Carr, D.~Massatt, S.~B. Torrisi, P.~Cazeaux, M.~Luskin, E.~Kaxiras,
  Relaxation and domain formation in incommensurate two-dimensional
  heterostructures, Phys. Rev. B 98 (2018) 224102.
\newblock \href {https://doi.org/10.1103/PhysRevB.98.224102}
  {\path{doi:10.1103/PhysRevB.98.224102}}.

\bibitem{Halbertal2021}
D.~Halbertal, N.~R. Finney, S.~S. Sunku, A.~Kerelsky, C.~Rubio-Verd{\'u},
  S.~Shabani, L.~Xian, S.~Carr, S.~Chen, C.~Zhang, L.~Wang,
  D.~Gonzalez-Acevedo, A.~S. McLeod, D.~Rhodes, K.~Watanabe, T.~Taniguchi,
  E.~Kaxiras, C.~R. Dean, J.~C. Hone, A.~N. Pasupathy, D.~M. Kennes, A.~Rubio,
  D.~N. Basov, Moir{\'e} metrology of energy landscapes in van der {W}aals
  heterostructures, Nature Communications 12 (2021) 242.
\newblock \href {https://doi.org/10.1038/s41467-020-20428-1}
  {\path{doi:10.1038/s41467-020-20428-1}}.

\bibitem{Kaliteevsk2023}
M.~A. Kaliteevski, V.~Enaldiev, V.~I. Fal’ko, Twirling and spontaneous
  symmetry breaking of domain wall networks in lattice-reconstructed
  heterostructures of two-dimensional materials, Nano Letters 23 (2023)
  8875--8880.
\newblock \href {https://doi.org/10.1021/acs.nanolett.3c01896}
  {\path{doi:10.1021/acs.nanolett.3c01896}}.

\bibitem{Brown2012}
L.~Brown, R.~Hovden, P.~Huang, M.~Wojcik, D.~A. Muller, J.~Park, Twinning and
  twisting of tri- and bilayer graphene, Nano Lett. 12 (2012) 1609--1615.
\newblock \href {https://doi.org/10.1021/nl204547v}
  {\path{doi:10.1021/nl204547v}}.

\bibitem{Baskin1955}
Y.~Baskin, L.~Meyer, Lattice constants of graphite at low temperatures, Phys.
  Rev. 100 (1955) 544--544.
\newblock \href {https://doi.org/10.1103/PhysRev.100.544}
  {\path{doi:10.1103/PhysRev.100.544}}.

\bibitem{Pierson1993}
H.~O. Pierson, Handbook of Carbon, Graphite, Diamond, and Fullerenes:
  Properties, Processing and Applications, Noyes Publications, Park Ridge, NJ,
  USA, 1993.

\bibitem{Zacharia2004}
R.~Zacharia, H.~Ulbricht, T.~Hertel, Interlayer cohesive energy of graphite
  from thermal desorption of polyaromatic hydrocarbons, Phys. Rev. B 69 (2004)
  155406.
\newblock \href {https://doi.org/10.1103/PhysRevB.69.155406}
  {\path{doi:10.1103/PhysRevB.69.155406}}.

\bibitem{Girifalco1956}
L.~A. Girifalco, R.~A. Lad, Energy of cohesion, compressibility, and the
  potential energy functions of the graphite system, J. Chem. Phys. 25 (1956)
  693--697.
\newblock \href {https://doi.org/10.1063/1.1743030}
  {\path{doi:10.1063/1.1743030}}.

\bibitem{Benedict1998}
L.~X. Benedict, N.~G. Chopra, M.~L. Cohen, A.~Zettl, S.~G. Louie, V.~H. Crespi,
  Microscopic determination of the interlayer binding energy in graphite, Chem.
  Phys. Lett. 286 (1998) 490--496.
\newblock \href {https://doi.org/10.1016/S0009-2614(97)01466-8}
  {\path{doi:10.1016/S0009-2614(97)01466-8}}.

\bibitem{Alden2013}
J.~S. Alden, A.~W. Tsen, P.~Y. Huang, R.~Hovden, L.~Brown, J.~Park, D.~A.
  Muller, P.~L. McEuen, Strain solitons and topological defects in bilayer
  graphene, PNAS 110 (2013) 11256--11260.
\newblock \href {https://doi.org/10.1073/pnas.1309394110}
  {\path{doi:10.1073/pnas.1309394110}}.

\bibitem{Boschetto2013}
D.~Boschetto, L.~Malard, C.~H. Lui, K.~F. Mak, Z.~Li, H.~Yan, T.~F. Heinz,
  Real-time observation of interlayer vibrations in bilayer and few-layer
  graphene, Nano Lett. 13 (2013) 4620--4623.
\newblock \href {https://doi.org/10.1021/nl401713h}
  {\path{doi:10.1021/nl401713h}}.

\bibitem{Tan2012}
P.~H. Tan, W.~P. Han, W.~J. Zhao, Z.~H. Wu, K.~Chang, H.~Wang, Y.~F. Wang,
  N.~Bonini, N.~Marzari, N.~Pugno, G.~Savini, A.~Lombardo, A.~C. Ferrari, The
  shear mode of multilayer graphene, Nature Materials 11 (2012) 294--300.
\newblock \href {https://doi.org/10.1038/nmat3245}
  {\path{doi:10.1038/nmat3245}}.

\bibitem{Giannozzi2020}
P.~Giannozzi, O.~Baseggio, P.~Bonf\`{a}, D.~Brunato, R.~Car, I.~Carnimeo,
  C.~Cavazzoni, S.~de~Gironcoli, P.~Delugas, F.~Ferrari~Ruffino, A.~Ferretti,
  N.~Marzari, I.~Timrov, A.~Urru, S.~Baroni, Quantum {ESPRESSO} toward the
  exascale, J. Chem. Phys. 152 (2020) 154105.
\newblock \href {https://doi.org/10.1063/5.0005082}
  {\path{doi:10.1063/5.0005082}}.

\bibitem{Giannozzi2017}
P.~Giannozzi, O.~Andreussi, T.~Brumme, O.~Bunau, M.~B. Nardelli, M.~Calandra,
  R.~Car, C.~Cavazzoni, D.~Ceresoli, M.~Cococcioni, N.~Colonna, I.~Carnimeo,
  A.~{Dal Corso}, S.~de~Gironcoli, P.~Delugas, R.~A. DiStasio, A.~Ferretti,
  A.~Floris, G.~Fratesi, G.~Fugallo, R.~Gebauer, U.~Gerstmann, F.~Giustino,
  T.~Gorni, J.~Jia, M.~Kawamura, H.-Y. Ko, A.~Kokalj,
  E.~K{\"u}\c{c}{\"u}kbenli, M.~Lazzeri, M.~Marsili, N.~Marzari, F.~Mauri,
  N.~L. Nguyen, H.-V. Nguyen, A.~{Otero-de-la-Roza}, L.~Paulatto, S.~Ponc\'{e},
  D.~Rocca, R.~Sabatini, B.~Santra, M.~Schlipf, A.~P. Seitsonen, A.~Smogunov,
  I.~Timrov, T.~Thonhauser, P.~Umari, N.~Vast, X.~Wu, S.~Baroni, Advanced
  capabilities for materials modelling with {Quantum ESPRESSO}, J. Phys.:
  Condens. Matter 29 (2017) 465901.
\newblock \href {https://doi.org/https://doi.org/10.1088/1361-648X/aa8f79}
  {\path{doi:https://doi.org/10.1088/1361-648X/aa8f79}}.

\bibitem{Giannozzi2009}
P.~Giannozzi, S.~Baroni, N.~Bonini, M.~Calandra, R.~Car, C.~Cavazzoni,
  D.~Ceresoli, G.~L. Chiarotti, M.~Cococcioni, I.~Dabo, A.~{Dal Corso},
  S.~de~Gironcoli, S.~Fabris, G.~Fratesi, R.~Gebauer, U.~Gerstmann,
  C.~Gougoussis, A.~Kokalj, M.~Lazzeri, L.~Martin-Samos, N.~Marzari, F.~Mauri,
  R.~Mazzarello, S.~Paolini, A.~Pasquarello, L.~Paulatto, C.~Sbraccia,
  S.~Scandolo, G.~Sclauzero, A.~P. Seitsonen, A.~Smogunov, P.~Umari, R.~M.
  Wentzcovitch, {QUANTUM ESPRESSO}: a modular and open-source software project
  for quantum simulations of materials, J. Phys.: Condens. Matter 21 (2009)
  395502.
\newblock \href {https://doi.org/10.1088/0953-8984/21/39/395502}
  {\path{doi:10.1088/0953-8984/21/39/395502}}.

\bibitem{QE}
Quantum ESPRESSO code: http://www.quantum-espresso.org for BLAS technical
  forum, 2020.

\bibitem{Chakraborty2020}
D.~Chakraborty, K.~Berland, T.~Thonhauser, Next-generation nonlocal van der
  {W}aals density functional, J. Chem. Theory and Computation 16 (2020)
  5893--5911.
\newblock \href {https://doi.org/10.1021/acs.jctc.0c00471}
  {\path{doi:10.1021/acs.jctc.0c00471}}.

\bibitem{Thonhauser2015}
T.~Thonhauser, S.~Zuluaga, C.~A. Arter, K.~Berland, E.~Schr\"oder,
  P.~Hyldgaard, Spin signature of nonlocal correlation binding in metal-organic
  frameworks, Phys. Rev. Lett. 115 (2015) 136402.
\newblock \href {https://doi.org/10.1103/PhysRevLett.115.136402}
  {\path{doi:10.1103/PhysRevLett.115.136402}}.

\bibitem{Thonhauser2007}
T.~Thonhauser, V.~R. Cooper, S.~Li, A.~Puzder, P.~Hyldgaard, D.~C. Langreth,
  Van der {W}aals density functional: Self-consistent potential and the nature
  of the van der {W}aals bond, Phys. Rev. B 76 (2007) 125112.
\newblock \href {https://doi.org/10.1103/PhysRevB.76.125112}
  {\path{doi:10.1103/PhysRevB.76.125112}}.

\bibitem{Berland2015}
K.~Berland, V.~R. Cooper, K.~Lee, E.~Schr\"oder, T.~Thonhauser, P.~Hyldgaard,
  B.~I. Lundqvist, Van der {W}aals forces in density functional theory: a
  review of the {vdW-DF} method, Reports on Progress in Physics 78 (2015)
  066501.
\newblock \href {https://doi.org/10.1088/0034-4885/78/6/066501}
  {\path{doi:10.1088/0034-4885/78/6/066501}}.

\bibitem{Langreth2009}
D.~C. Langreth, B.~I. Lundqvist, S.~D. Chakarova-K\"{a}ck, V.~R. Cooper,
  M.~Dion, P.~Hyldgaard, A.~Kelkkanen, J.~Kleis, L.~Kong, S.~Li, P.~G. Moses,
  E.~Murray, A.~Puzder, H.~Rydberg, E.~Schr\"{o}der, T.~Thonhauser, A density
  functional for sparse matter, Journal of Physics: Condensed Matter 21 (2009)
  084203.
\newblock \href {https://doi.org/10.1088/0953-8984/21/8/084203}
  {\path{doi:10.1088/0953-8984/21/8/084203}}.

\bibitem{Lebedeva2016a}
I.~V. Lebedeva, A.~V. Lebedev, A.~M. Popov, A.~A. Knizhnik, Comparison of
  performance of van der {W}aals-corrected exchange-correlation functionals for
  interlayer interaction in graphene and hexagonal boron nitride, Comput.
  Mater. Sci. 128 (2017) 45--58.
\newblock \href {https://doi.org/10.1016/j.commatsci.2016.11.011}
  {\path{doi:10.1016/j.commatsci.2016.11.011}}.

\bibitem{Hamann2013}
D.~R. Hamann, Optimized norm-conserving {V}anderbilt pseudopotentials, Phys.
  Rev. B 88 (2013) 085117.
\newblock \href {https://doi.org/10.1103/PhysRevB.88.085117}
  {\path{doi:10.1103/PhysRevB.88.085117}}.

\bibitem{Hamann2017}
D.~R. Hamann, Erratum: {O}ptimized norm-conserving {V}anderbilt
  pseudopotentials [{Phys. Rev. B 88 (2013) 085117}], Phys. Rev. B 95 (2017)
  239906.
\newblock \href {https://doi.org/10.1103/PhysRevB.95.239906}
  {\path{doi:10.1103/PhysRevB.95.239906}}.

\bibitem{Setten2018}
M.~J. {van Setten}, M.~Giantomassi, E.~Bousquet, M.~J. Verstraete, D.~R.
  Hamann, X.~Gonze, G.-M. Rignanese, The {PseudoDojo}: Training and grading a
  85 element optimized norm-conserving pseudopotential table, Computer Physics
  Communications 226 (2018) 39--54.
\newblock \href {https://doi.org/https://doi.org/10.1016/j.cpc.2018.01.012}
  {\path{doi:https://doi.org/10.1016/j.cpc.2018.01.012}}.

\bibitem{pseudo-dojo}
{Pseudo Dojo} database: https://www.pseudo-dojo.org/, accessed 2025.

\bibitem{Greshnyakov2019}
V.~A. Greshnyakov, E.~A. Belenkov, M.~M. Brzhezinskaya, Theoretical
  investigation of phase transitions of graphite and cubic {3C} diamond into
  hexagonal {2H} diamond under high pressures, Phys. Status Solidi B 256 (2019)
  1800575.
\newblock \href {https://doi.org/10.1002/pssb.201800575}
  {\path{doi:10.1002/pssb.201800575}}.

\bibitem{Popov2011}
A.~M. Popov, I.~V. Lebedeva, A.~A. Knizhnik, Y.~E. Lozovik, B.~V. Potapkin,
  Commensurate-incommensurate phase transition in bilayer graphene, Phys. Rev.
  B 84 (2011) 045404.
\newblock \href {https://doi.org/10.1103/PhysRevB.84.045404}
  {\path{doi:10.1103/PhysRevB.84.045404}}.

\bibitem{Monkhorst1976}
H.~J. Monkhorst, J.~D. Pack, Special points for {B}rillouin-zone integrations,
  Phys. Rev. B 13 (1976) 5188--5192.
\newblock \href {https://doi.org/10.1103/PhysRevB.13.5188}
  {\path{doi:10.1103/PhysRevB.13.5188}}.

\bibitem{Brenner2002}
D.~W. Brenner, O.~A. Shenderova, J.~A. Harrison, S.~J. Stuart, B.~Ni, S.~B.
  Sinnott, A second-generation reactive empirical bond order ({REBO}) potential
  energy expression for hydrocarbons, J. Phys.: Condens. Matter 14 (2002)
  783--802.
\newblock \href {https://doi.org/10.1088/0953-8984/14/4/312}
  {\path{doi:10.1088/0953-8984/14/4/312}}.

\bibitem{Surinlert2022}
P.~Surinlert, P.~Kokmat, A.~Ruammaitree, Growth of turbostratic stacked
  graphene using waste ferric chloride solution as a feedstock, RSC Adv. 12
  (2022) 25048--25053.
\newblock \href {https://doi.org/10.1039/D2RA02686D}
  {\path{doi:10.1039/D2RA02686D}}.

\bibitem{Xu2021}
Z.~Xu, S.~Nakamura, T.~Inoue, Y.~Nishina, Y.~Kobayashi, Bulk-scale synthesis of
  randomly stacked graphene with high crystallinity, Carbon 185 (2021)
  368--375.
\newblock \href {https://doi.org/10.1016/j.carbon.2021.09.034}
  {\path{doi:10.1016/j.carbon.2021.09.034}}.

\bibitem{Kokmat2023}
P.~Kokmat, P.~Surinlert, A.~Ruammaitree, Growth of high-purity and high-quality
  turbostratic graphene with different interlayer spacings, ACS Omega 8 (2023)
  4010--4018.
\newblock \href {https://doi.org/10.1021/acsomega.2c06834}
  {\path{doi:10.1021/acsomega.2c06834}}.

\bibitem{Perdew1996}
J.~P. Perdew, K.~Burke, M.~Ernzerhof, Generalized gradient approximation made
  simple, Phys. Rev. Lett. 77 (1996) 3865--3868.
\newblock \href {https://doi.org/10.1103/PhysRevLett.77.3865}
  {\path{doi:10.1103/PhysRevLett.77.3865}}.

\bibitem{Pokrovskii1978}
V.~L. Pokrovski\u{i}, A.~L. Talapov, Phase transitions and vibrational spectra
  of almost commensurate structures, Sov. Phys. J. Exp. Theor. Phys. 48 (1978)
  579--582.

\bibitem{Kumar2016}
H.~Kumar, L.~Dong, V.~B. Shenoy, Limits of coherency and strain transfer in
  flexible {2D} van der {W}aals heterostructures: Formation of strain solitons
  and interlayer debonding, Scientific Reports 6 (2016) 21516.
\newblock \href {https://doi.org/10.1038/srep21516}
  {\path{doi:10.1038/srep21516}}.

\end{thebibliography}
\end{document}